\documentclass[prl,twocolumn,altaffilletter,nolongbibliography,numerical,flushbottom,secnumarabic,superscriptaddress,floatfix,nobalancelastpage]{revtex4-1}

\usepackage{graphicx}
\usepackage{braket}
\usepackage{amsmath, amssymb}
\usepackage{booktabs}  
\usepackage{comment}
\usepackage[normalem]{ulem}

\usepackage[breaklinks, pdftex, hyperfootnotes=true, pdfpagelabels, bookmarks, pageanchor]{hyperref}
\hypersetup{%
	colorlinks=true, linktocpage=true, pdfstartpage=1, pdfstartview=FitH, pdfborder={0 0 0},%
	breaklinks=true, pdfpagemode=UseNone, pageanchor=true, pdfpagemode=UseOutlines,%
	plainpages=false, bookmarksnumbered, bookmarksopen=true, bookmarksopenlevel=1,%
	hypertexnames=true, pdfhighlight=/O,
	urlcolor=red, linkcolor=red, citecolor=red,
	}
\usepackage{orcidlink}

\newcommand{\xt}[1]{ #1}
\newcommand{\blue}[1]{ #1}

\usepackage{soul}
\newcommand{\add}[1]{#1 }
\newcommand{\del}[1]{}

\begin{document}

\title{Many-body Dynamics in Monitored Atomic Gases Without Post-Selection Barrier }

\author{Gianluca Passarelli~\orcidlink{0000-0002-3292-0034}}
\thanks{These two authors contributed equally to the work}
\affiliation{CNR-SPIN, c/o Complesso di Monte S. Angelo, via Cinthia - 80126 - Napoli, Italy}

\author{Xhek Turkeshi~\orcidlink{0000-0003-1093-3771}}
\thanks{These two authors contributed equally to the work}
\affiliation{JEIP, UAR 3573 CNRS, Coll\`ege de France, PSL Research University, 75321 Paris Cedex 05, France}

\author{Angelo Russomanno}
\affiliation{Scuola Superiore Meridionale, Universit\`a di Napoli Federico II, I-80138 Napoli, Italy}

\author{Procolo Lucignano}
\affiliation{Dipartimento di Fisica, Universit\`a di Napoli Federico II, I-80126 Napoli, Italy}

\author{Marco Schir\`o}
\affiliation{JEIP, UAR 3573 CNRS, Coll\`ege de France, PSL Research University, 75321 Paris Cedex 05, France}
 
\author{Rosario Fazio~\orcidlink{0000-0002-7793-179X}}
\affiliation{The Abdus Salam International Center for Theoretical Physics,  34151 Trieste, Italy}
\affiliation{Dipartimento di Fisica, Universit\`a di Napoli Federico II, I-80126 Napoli, Italy}

\date{\today} 

\begin{abstract}
We study the properties of a monitored ensemble of atoms driven by a laser field and in the presence of collective 
decay. The properties of the quantum trajectories describing the atomic cloud drastically depend on the monitoring protocol and are distinct from those of the average density matrix. 
By varying the strength of the external drive, a measurement-induced phase transition occurs separating two phases with entanglement entropy scaling sub-extensively with the system size. Incidentally, the critical point coincides with the superradiance transition of the trajectory-averaged dynamics.
Our setup is implementable in current light-matter
interaction devices, and most notably, the monitored dynamics is free from
the post-selection measurement problem, even in the case of imperfect
monitoring.
\end{abstract}

\maketitle

\textit{Introduction ---}  When a quantum system is externally measured, its dynamic is strongly altered. The decay of 
an atom, for example, will occur through a sudden quantum jump from the excited to the ground state at a random time.  The evolution 
of quantum systems along chosen trajectories has been intensively investigated for almost forty 
years~\cite{carmichael1999statisticalmethodsin,wiseman2009quantummeasurementand,jacobs2014quantummeasurementtheory}. Only very recently, however,  monitored dynamics entered the world of many-body 
systems. Two independent works~\cite{li2018quantumzenoeffect,skinner2019measurementinducedphase} showed that a quantum many-body system subject to a mixed 
evolution composed of unitary intervals interrupted by local measurements undergoes a transition in its 
quantum correlation pattern
which is invisible to the properties of the average density matrix. 
Only by resolving the dynamics along each trajectory it is possible to construct non-linear functions of quantum states, such as entanglement measures or trajectory correlations, able to reveal these so-called Measurement-Induced Phase Transitions (MIPT). 
An extensive research activity followed these initial works~\cite{fisher2023randomquantumcircuits,potter2022quantumsciencesandtechnology,lunt2022quantumsciencesandtechnology}, analyzing salient aspects of measurement-induced phases in monitored quantum 
circuits~\cite{li2019measurementdrivenentanglement,szyniszewski2019entanglementtransitionfrom,jian2020measurementinducedcriticality,li2021statisticalmechanicsmodel,zabalo202criticalpropertiesof,szyniszewski2020universalityofentanglement,turkeshi2020measurementinducedcriticality,lunt2021measurementinducedcriticality,sierant2022measurementinducedphase,nahum2021measurementandentanglement,zabalo2022operatorscalingdimensions,sierant2022universalbehaviorbeyond,chiriaco2023diagrammaticmethodfor,klocke2023majorana}, non-interacting~\cite{cao2019entanglementina,nahum2020entanglementanddynamics,buchhold2021effectivetheoryfor,chaoming2022criticalityandentanglement,coppola2022growthofentanglement,fava2023nonlinearsigmamodels,poboiko2023theoryoffree,jian2023measurementinducedentanglement,merritt2023entanglementtransitionswith,alberton2021entanglementtransitionin,turkeshi2021measurementinducedentanglement,turkeshi2022entanglementtransitionsfrom,piccitto2022entanglementtransitionsin,piccitto2023entanglementdynamicswith,tirrito2023fullcountingstatistics,paviglianiti2023multipartiteentanglementin} and interacting~\cite{rossini2020measuremendinduceddynamics,tang2020measurementinducedphase,fuji2020measurementinducedquantum,sierant2022dissipativefloquetdynamics,dogger2022generalizedquantummeasurements,altland2022dynamicsofmeasured} monitored Hamiltonian systems. 
Common to these framework is the variety of entanglement patterns induced by measurements, that are deeply tied to the encoding/decoding properties of quantum channels~\cite{gullans2020scalableprobesof,gullans2020dynamicalpurificationphase,loio2023purificationtimescalesin,choi2020quantumerrorcorrection,bao2020theoryofthe,bao2021symmetryenrichedphases,fidkowski2021howdynamicalquantum,bao2021finitetimeteleportation,barratt2022transitionsinthe,dehgani2022neuralnetworkdecoders,kelly2022coherencerequirementsfor}.

\begin{figure*}[t!]
    \centering
    \includegraphics[width=\textwidth]{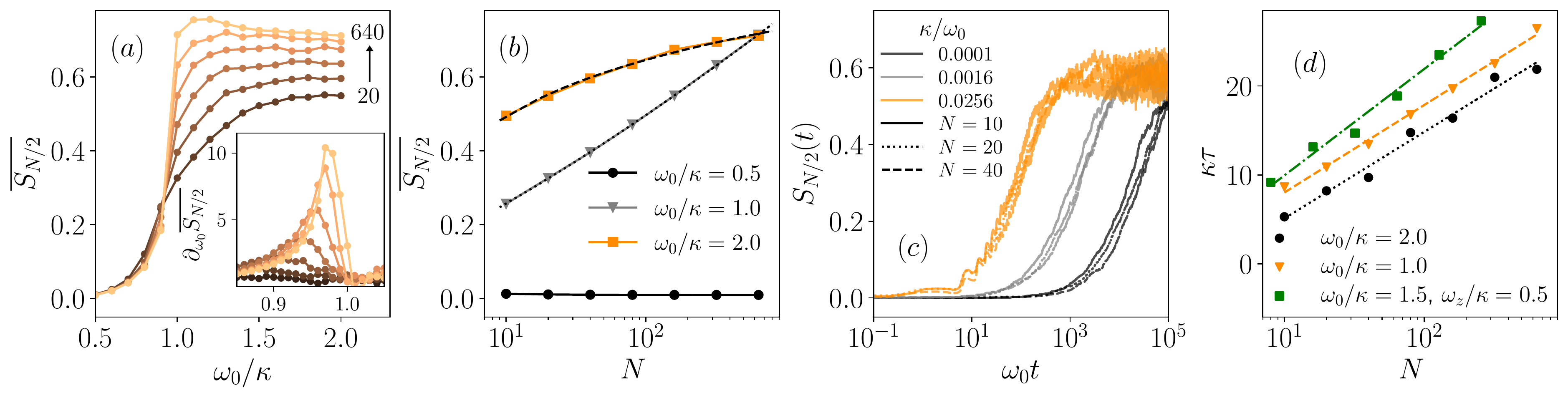}
    \caption{Results of QJ simulations. (a)~Long-time average of the half-chain entanglement entropy $\overline{S_{N/2}}$ as a function of $\omega_0 / \kappa$, for system sizes $N = 20, 40, 80, 160, 320, 640$. In the inset, the numerical derivative of the long-time averaged entanglement entropy highlights the critical point; the peak height 
    grows logarithmically with $N$. (b)~Scaling of $\overline{S_{N/2}}$ as a function of $N$ at fixed values of $\omega_0 / \kappa$. At the critical point,  
    the long-time averaged entanglement entropy grows logarithmically with $N$ (gray triangles). In the phase $\omega_0 / \kappa < 1$ the behavior is 
    area-law (black circles). The scaling in the region $\omega_0 / \kappa > 1$ is sub-logarithmic (orange squares). Numerically, one can fit the curves as 
    $\overline{S_{N/2}} \sim  \ln^\beta N$ with a non-universal exponent $\beta < 1$ ($\beta \sim 0.2$ at $\omega_0 / \kappa =2$). The exponent 
    decreases when moving away from the critical point. (c)~Long-time dynamics of the entanglement entropy for several system sizes and several values of $\kappa/\omega_0$.
    The entanglement entropy grows as $\ln t$ before saturating.  (d)~Dependence of the saturation time on the system size $N$ (in all cases, $\tau \sim \ln N$).}
    \label{fig:qj}
\end{figure*}

Despite this large theoretical effort, the experimental evidence of MIPTs is much more limited with only three pioneering experimental works at present~\cite{noel2021measurementinducedquantum,koh2022experimentalrealizationof,hoke2023quantuminformationphases}. 
Following a first quantum simulation with trapped ions~\cite{noel2021measurementinducedquantum} the scaling close to the critical point has been explored with the IBM quantum  processor~\cite{koh2022experimentalrealizationof}. 
A fundamental reason hinders the possibility of observing monitored phases, known as the post-selection problem. 
To perform averages of observables along a given trajectory, one should reproduce the same sequence of random jumps with sufficient probability. 
This task is challenging as the probability of reproducing the same trajectory scales to zero exponentially with system size and timescale, explaining why experiments were limited to a few sites and the considerable efforts required to increase the lattice length.  
The post-selection barrier can be avoided by quantum-classical approaches that combine measurement outcomes to classical post-processing~\cite{gullans2020scalableprobesof,li2022cross,li2021robustdecodingin,garratt2023probing}. These methods however require a perfect detection and the ability to efficiently simulate the quantum dynamics on a classical computer.  
In certain quantum circuits, post-selection overhead can be mitigated by resorting to space-time duality~\cite{ippoliti2021postselectionfreeentanglement,lu2021spacetimeduality,hoke2023quantuminformationphases}. 
Another approach was proposed in adaptive quantum circuits~\cite{iadecola2022dynamicalentanglementtransition,buchhold2022revealingmeasurementinduced}, where the measurement registry conditions the evolution so to render MIPTs visible in the average dynamics (at density matrix level). However, this route generally fails as feedback may alter the physics of the system, and separate transitions may occur in the monitored and average dynamics~\cite{ravindranath2022entanglementsteeringin,odea2022entanglementandabsorbing,piroli2022trivialityofquantum,sierant2023controllingentanglementat}.
The post-selection problem remains a formidable hurdle to overcome, and the search for cases where it can be mitigated is necessary for experimental progress in monitoring quantum many-body systems. 

Here we show that there is a class of infinite-range spin systems where monitored many-body dynamics can be efficiently realized with a post-selection overhead growing, at most, \emph{polynomially} with the system size. These systems have a non-trivial monitored dynamics, 
which is yet experimentally accessible in atomic ensembles driven by a laser field and in the presence of a collective decay. Here, the type of measurements is pivotal in determining the entanglement behavior of the quantum trajectories describing the system. Different monitoring protocols lead to a different scaling of entanglement measures throughout the phase diagram, highlighting the measurement-induced nature of the system.
The quantum trajectories describing the system undergo a MIPT separating two sub-volume law behaviors.
Incidentally, this measurement-induced criticality coincide with the dissipative (in this case, superradiant) transition of the average dynamics, 
a possibility already discussed in the literature~\cite{sierant2022dissipativefloquetdynamics,ravindranath2022entanglementsteeringin,odea2022entanglementandabsorbing,piroli2022trivialityofquantum,sierant2023controllingentanglementat}. As in these cases, critical and off-critical features of the dissipative and monitoring-induced phases are generally inequivalent, as the mechanism underpinning these phenomena is different. \blue{(We have further discussed this fact in~\cite{supplementary} with a concrete example having a phase transition at the average level, but lacking a MIPT.)}.

The system consists of a cloud of $N$ atoms (each behaving as a two-level system with associated Pauli matrices $ \hat{\sigma}_i^\alpha$, $\alpha = x, y, z$ for the $i$-th atom), driven by an external laser with collective decay~\cite{ferioli2023anonequilibrium}. 
In the absence of monitoring, its density matrix obeys  the Lindblad equation
\begin{equation}
\label{eq:btc}
    \dot{\hat\rho} = \mathcal{L}(\hat\rho)\equiv -i [\hat{\cal H}, \hat\rho] + \frac{\kappa}{J} \Bigl(\hat{J}_- \hat\rho \hat{J}_+ - \frac{1}{2}\lbrace \hat{J}_+ \hat{J}_-, \hat{\rho}\rbrace \Bigr),
\end{equation}
where ${\hat{J}_\alpha = \sum_i \hat{\sigma}_i^\alpha/2}$,  ${\hat{J}_\pm = \hat{J}_x \pm i \,\hat{J}_y}$, ${J = N/2}$ is the total spin,  and $  {\hat{\cal  H} = \omega_0 \hat{J}_x }$. 
The steady state has a superradiant phase transition separating a normal from a time-crystal phase~\cite{iemini2018boundarytimecrystals,hannukainen2018dissipationdrivenquantum,carollo:thermodynamics-btc}. 
The dynamic governed by~\eqref{eq:btc} was recently realized experimentally~\cite{ferioli2023anonequilibrium}. 
Following the evolution along a quantum trajectory, \textit{i.\,e.} unraveling~\eqref{eq:btc}, requires specifying a monitoring protocol~\cite{cabot2022quantum,poggi2023measurementinducedmmultipartite}.
As we detail below, despite the average dynamics is the same [cf.~\eqref{eq:btc}], different types of measurements generate inequivalent trajectory ensembles, \blue{a fact already noted in, \emph{e.g.} ~\cite{alberton2021entanglementtransitionin,turkeshi2021measurementinducedentanglement,piccitto2022entanglementtransitionsin}}.
\blue{Their distinct} traits are showcased in any non-linear function of the state, \emph{e.g.} in entanglement measures. We will separately discuss two, experimentally motivated, monitoring protocols. 
The system is either coupled to a photodetector, where measurement act as quantum jumps (QJ), or to a homodyne detector, continuously probing the system and lead to a quantum state diffusion (QSD).  We quantify the entanglement in the fluctuating steady state through entanglement entropy and  quantum Fisher information.

\textit{Quantum jumps ---} In this case, the system evolves according to a (smooth) effective non-Hermitian Hamiltonian 
$\hat{\mathcal{H}}_\text{nj} = \hat{\mathcal{H}} - i\,(\kappa/2J) \hat{J}_+ \hat{J}_-$, interrupted, at random times, by quantum jumps when the wavefunction $\ket{\psi (t)}$ changes abruptly as 
\begin{equation}
    \ket{\psi (t_+)} = \frac{\hat{J}_- \ket{\psi (t_-)}}{\sqrt{\bra{\psi (t_-)} \hat{J}_+ \hat{J}_- \ket{\psi (t_-)}}}.\label{eq:qj}
\end{equation}
In a time-interval  $\delta t$, jumps occur with a probability  $\delta p = \kappa \delta t \langle \hat{J}_+ \hat{J}_- \rangle / J $. Details of the numerics are 
given in~\cite{supplementary}.

 \begin{figure*}[t]
    \centering
    \includegraphics[width=\textwidth]{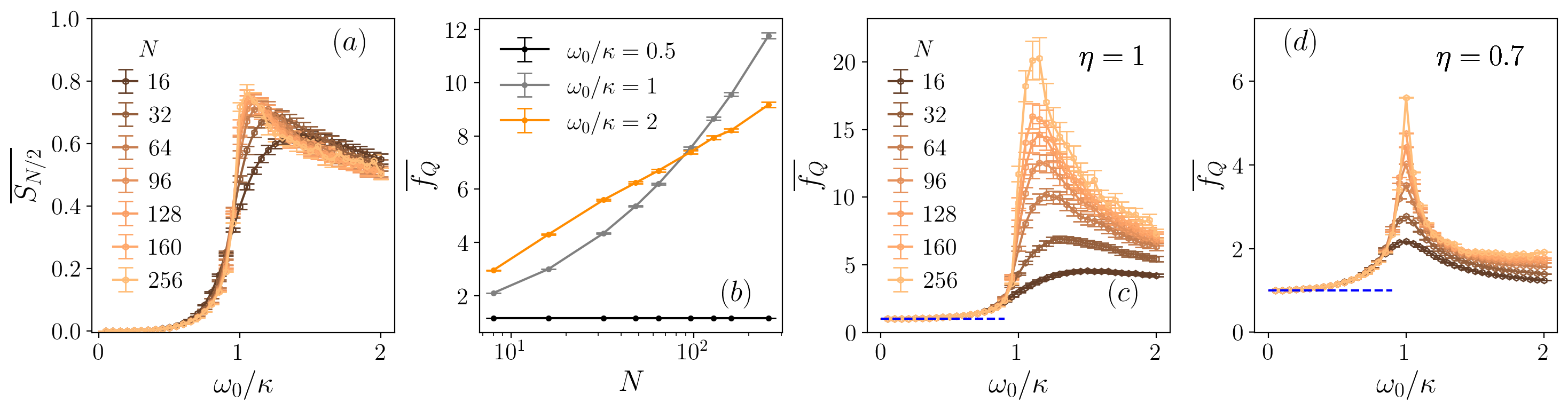}
    \caption{Results of QSD simulations. (a)~Long-time averaged entanglement entropy $\overline{S_{N/2}}$ as a function of $\omega_0 / \kappa$ for various system sizes $N$. The entanglement develops a peak at the critical point $\omega_0 = \kappa$. 
    (b)~Scaling of $\overline{f_Q}=\overline{F_Q}/N$ as a function of $N$ at the fixed values of $\omega_0 / \kappa$. 
    At the critical point, the long-time averaged Fisher density grows $\propto\sqrt{N}$ (grey line). In the phase $\omega_0< \kappa$, $\overline{f_Q}$ saturates to the constant $1$ (black line). For $\omega_0>\kappa$, the Fisher density scales logarithmically with the system size. Therefore, the system is multipartite entangled at the critical point and in the non-trivial phase. 
    As discussed in the text, we compare the Fisher density for (c) efficient ($\eta=0$) and (d) inefficient detectors ($\eta=0.7$). The blue-dashed line is the constant value $\overline{f_Q} = 1$. While quantitative changes are present, the qualitative features of the phase diagram are preserved.}
    \label{fig:qsd_fit}
\end{figure*}

The QJ unraveling leads to a ensemble of stochastic trajectories fixed by the occurrence (time) of the jumps. The Lindblad evolution~\eqref{eq:btc} describes the average features of this ensemble~\cite{wiseman2009quantummeasurementand}, but does not capture its higher-cumulants. These features are probed by non-linear functions of the state, \emph{e.g.}, the entanglement measures considered in this manuscript~\footnote{This fact hold because the trajectory average does not commute with non-linear functions of the quantum trajectory~\cite{cao2019entanglementina}.}.

\xt{We first consider the entanglement entropy, defined for a pure state $\ket{\psi (t)}$ and a bipartition of the system into two sets $A$ and $B$, by $S_{A}(\ket{\psi (t)} ) = -\mbox{Tr}_A(\hat{\rho}_A \ln \hat{\rho}_A)$. Here, $\hat{\rho}_A = \mbox{Tr}_B \ket{\psi (t)} \bra{\psi (t)}$ is the partial trace over the degrees of freedom of subsystem $B$.}
We will consider balanced partitions with $N_A=N_B=N/2$, and the entanglement entropy at long times is then averaged over the quantum trajectories and over the time domain $\overline{S_{N/2}}\;$.

In the evolution dictated by~\eqref{eq:qj}, quantum correlations are built by collective jumps, the external drive leads to phase shifts between different components of the quantum state. Their interplay leads to a non-trivial behaviour of the entanglement for quantum trajectories. 
In Fig.~\ref{fig:qj}(a), we plot the averaged entropy $\overline{S_{N/2}}$ as a function of $\omega_0 / \kappa$ for several values of $N$. 
In the small $\omega_0$ regime, the entanglement entropy is essentially independent of $N$: The system is in an area-law phase. 
An anomaly develops near $\omega_0 / \kappa \sim 1$. The singularity is clearly visible in the inset of Fig.~\ref{fig:qj}(a). 
The derivative of $\overline{S_{N/2}}$ has a logarithmic divergence of the peak height with $N$. 
The MIPT coincides with the superradiant transition of the average state, separating a normal from a time-crystal phase~\cite{iemini2018boundarytimecrystals}. This incidental fact~\cite{footnote1},
will \emph{not} be used in the following, as the average density matrix does not reveal the entanglement properties of the trajectory. 
At the critical point, the entanglement diverges logarithmically as shown in Fig.~\ref{fig:qj}(b).
For ${\omega_0 / \kappa > 1}$ the entropy grows more slowly with $N$. 
From the numerics, a good fit is obtained with  $\overline{S_{N/2}} \sim  \ln^\beta N$ with a non-universal exponent $\beta < 1$ that decreases moving away from the transition. In the limit $\omega_0 / \kappa \gg 1$, $\beta$ tends to zero. 
A more careful inspection based on of Fig.~\ref{fig:qj}(c,d) suggests $
\overline{S_{N/2}} \sim \ln \ln N $. The entropy grows as $\ln t$ up to a saturation time that depends logarithmically on $N$.
Sub-extensive phases in long-range circuits have been discussed in~\cite{block2022measurementinducedtransition,sharma2022measurementinducedcriticality,minato2022fateofmeasurementinduced,muller2022measurementinduceddark,hashizum2022measurementinducedphase,zhang2022universalentanglementtransitions}.

It is important to estimate the overhead that we should expect from the unavoidable post-selection in a brute-force experiment. The system analyzed here is free from the post-selection problem.
Because of the collective nature of the jumps, a quantum trajectory can be represented as a binary string with a record of the sequence of jumps (once a time-bin has been fixed). 
The probability of generating the same trajectory thus scales as $2^{-\tau}$ with $\tau$ of the order of the saturation time. 
For ${\omega_0 / \kappa < 1}$, $\tau$ is independent on $N$~\cite{supplementary}. In the sub-extensive regime ${\omega_0 / \kappa > 1}$ and at the critical point ${\omega_0 / \kappa = 1}$, the saturation time grows logarithmically with the system size,
\begin{equation}
    \tau \sim a\log N + b,
\label{saturation}
\end{equation}
see Fig.~\ref{fig:qj}(d)
($a,b$ constants depending on $\omega_0 / \kappa $). The logarithmic scaling of the saturation time is a signature of collective dissipation~\cite{gross1982superradiance}, and can be estimated from the imaginary part of the non-Hermitian Hamiltonian~\cite{supplementary}.


Combining the behaviour of $\tau$ discussed above, it is straightforward to conclude that the probability of occurrence of a given trajectory is, at most, decaying as power-law $N^{-\gamma}$ (with $\gamma$ weakly dependent on the coupling constants).
The monitored dynamic of the system we consider is thus free from the post-selection problem. 

This property holds for a class of long-range interacting spin-systems with collective decay. We added an all-to-all term to the Hamiltonian of the form ${\cal H}_{z} = \omega_z J^2_z$ and the scaling of the saturation time, shown in Fig.~\ref{fig:qj}(d) green squares, is still logarithmic with the system size. We also considered the case of power-law decay exchange interaction among the spins. In this case, due to the absence of permutational invariance, we can consider much more modest system sizes. The results are reported in~\cite{supplementary}: as long as the interaction is sufficiently long-range the dynamics remains post-selection-free up to long times.


\textit{Quantum state diffusion ---} 
After discussing the quantum jumps evolution induced in the atomic cloud by the photodetector, we now consider the \emph{inequivalent} quantum state diffusion dynamics induced by a (non-ideal) homodyne detector~\cite{gisin1992thequantumstate,jacobs2014quantummeasurementtheory}. 
The dynamics is given by~\cite{wiseman2009quantummeasurementand}
\begin{equation}
	d\hat{\rho}_\mathrm{w} = \mathcal{L}(\hat{\rho}_\mathrm{w}) + \sqrt{\frac{\kappa \eta}{J}}dW ( \hat{J}_- \hat{\rho}_\mathrm{w} + \hat{\rho}_\mathrm{w} \hat{J}_+ - 2\langle \hat{J}_x\rangle \hat{\rho}_\mathrm{w} )\label{eq:qsd}
\end{equation}
where $dW$ is a Gaussian \^Ito noise with $\overline{dW}=0$ and $dW^2 = dt$. In Eq.~\eqref{eq:qsd} we have introduced the detector efficiency $\eta\in[0,1]$ to treat the effect of noise: $\eta=1$ corresponds to a perfect detector and $\eta<1$ describe imperfect detection, with the limit $\eta=0$ implying no trajectory resolution. 
Eq.~\eqref{eq:qsd} provides a different unraveling of the Lindblad equation in Eq.~\eqref{eq:btc} since $\hat\rho = \overline{\hat{\rho}_\mathrm{w}}$, and describes a system coupled to a homodyne detector, a framework of experimental relevance in current platforms~\cite{ferioliprivate}.
Crucially, the homodyne current 
\begin{equation}
    dI_t = \frac{\kappa}{J} \langle \hat{J}_x\rangle + \sqrt{\frac{\kappa \eta}{J}}dW \label{eq:dit}
\end{equation}
is experimentally detectable, thus allowing for an efficient monitoring (see below).
When $\eta=1$ the monitoring is perfect, the purity of an initial state is preserved in the dynamic. 
In this case, the entanglement entropy is a sensitive measure of quantum correlations in the system. In Fig.~\ref{fig:qsd_fit}(a) we present the average entanglement entropy varying $\omega_0/\kappa$ and for various system sizes. 
For $\omega_0<\kappa$ the entanglement behaves qualitatively as for the quantum jump evolution [cf.~Fig.~\ref{fig:qj}(a)]. 
Differently, for $\omega_0>\kappa$ the available system sizes show a saturation, hence an area-law phase. 
This is not surprising: the quantum state diffusion is a different unraveling of the Lindblad equation Eq.~\eqref{eq:btc} that is more prone to destroy entanglement, \textit{i.\,e.}, the infinite click limit of the quantum jump evolution. 
Remarkably, the phase transition occurs still at $\omega_0=\kappa$, where $\overline{S_{N/2}}$ develops a peak. 

Realistic experiments have $\eta<1$ due to decoherence to the environment, and Eq.~\eqref{eq:qsd} leads to mixed states, potentially altering the entanglement properties of the state~\cite{minoguchi2022continuousgaussianmeasurements,ladewig2022monitoredopenfermion,turkeshi2022enhancedentanglelmentnegativity,weinstein2022measurementinducedpower,weinstein2022scramblingtransitionin}. 
For instance, the von Neumann entropy is not an entanglement measure in this situation, as it also embodies classical correlations.
Therefore, to compare perfect and imperfect detectors, we study the quantum Fisher information (QFI), a measure of multipartite entanglement valid for pure and mixed states~\cite{Hauke2016,pezze2018quantummetrologywith,pappalardi2017multipartiteentanglementafter,pappalardi2022extensivemultipartiteentanglement,pappalardi2018scramblingandentanglement,brenes2020multipartitenetanglementstructure,dooley2023entanglementenhanced}.
The quantum Fisher information is $F_Q(\hat{\rho}_\mathrm{w}) = \max\mathrm{eigval}(M)\label{eq:qfidef}$ 
where the matrix $M_{\alpha,\beta} = 2 \sum_{k,l} \Lambda_{k,l}\langle k|\hat{J}_\alpha| l\rangle \langle l| \hat{J}_\beta| k\rangle$ is defined for the decomposition $\hat{\rho}_\mathrm{w} =\sum_{k} \lambda_k |k\rangle\langle k|$ and $\Lambda_{k,l}= (\lambda_k-\lambda_l)^2/(\lambda_k+\lambda_l)$~\footnote{\textit{En passant}, we note that for a pure state $\rho_\mathrm{w}=|\Psi\rangle_\mathrm{w}\langle\Psi|_\mathrm{w}$, $F_Q(\rho_\mathrm{w})$ is the maximal eigenvalue of the covariance matrix $M^\mathrm{cov}_{\alpha,\beta}=2 \langle \Psi_\mathrm{w} |\{ \hat{J}_\alpha,\hat{J}_\beta\} |\Psi\rangle- 4\langle \Psi_\mathrm{w} |\hat{J}_\alpha|\Psi_\mathrm{w}\rangle\langle \Psi_\mathrm{w} |\hat{J}_\beta|\Psi_\mathrm{w}\rangle$.}.
The QFI gives a bound to the multipartite entanglement entropy: if the Fisher density $f_Q\equiv F_Q/N$ is (strictly) larger than some divider $k$ of $L$, then the state contains entanglement between $(k+1)$ parties. The QFI is non-linear in the density matrix, hence $\overline{F_Q(\hat{\rho}_\mathrm{w})}\neq F_Q(\overline{\hat{\rho}_\mathrm{w}})$.
We first discuss the average QFI density $\overline{f_Q}$ in the limit $\eta=1$. 
In Fig.~\ref{fig:qsd_fit}(b) we show the system size scaling of $\overline{f_Q}$ in the normal phase, critical point, and boundary time crystal phase. 
In the former, the Fisher density saturates to the constant, signaling that the system is close to a product state. For $\omega_0>\kappa$ the Fisher density presents a logarithmic scaling with $N$~\cite{supplementary}. It follows that this phase exhibits area-law entanglement and logarithmic multipartite entanglement~\footnote{We note that in the quantum jump case, the Fisher density is extensive in the system size $\overline{f_Q}\sim N$, cf.~Ref.~\cite{supplementary}}.
At the critical point, a polynomial fit shows that $\overline{f_Q}\sim N^{1/2}$ ~\footnote{This fact reveals enhanced multipartite entanglement at the critical point}. 
In Fig.~\ref{fig:qsd_fit}(c) we summarize the behavior of the Fisher density for a perfect detector ($\eta=1$). 
The salient points of the above discussion are robust against noisy contribution. In Fig.~\ref{fig:qsd_fit}(d) we show that the phases and the peak of $\overline{f_Q}$ at the critical point are qualitatively unaltered for efficiency rate $\eta=0.7$. 
Further decreasing $\eta$ we would get close to the Lindblad framework Eq.~\eqref{eq:btc}, where the QFI has been computed in~\cite{fernando,fernando2}. 
The analysis presented above shows that experimental detection is feasible in the atomic systems of interest. In the next section, we discuss in more details an experimental implementation based on driven atomic gases and one on homodyne detection.

\begin{figure}[t]
    \centering
    \includegraphics[width=\columnwidth]{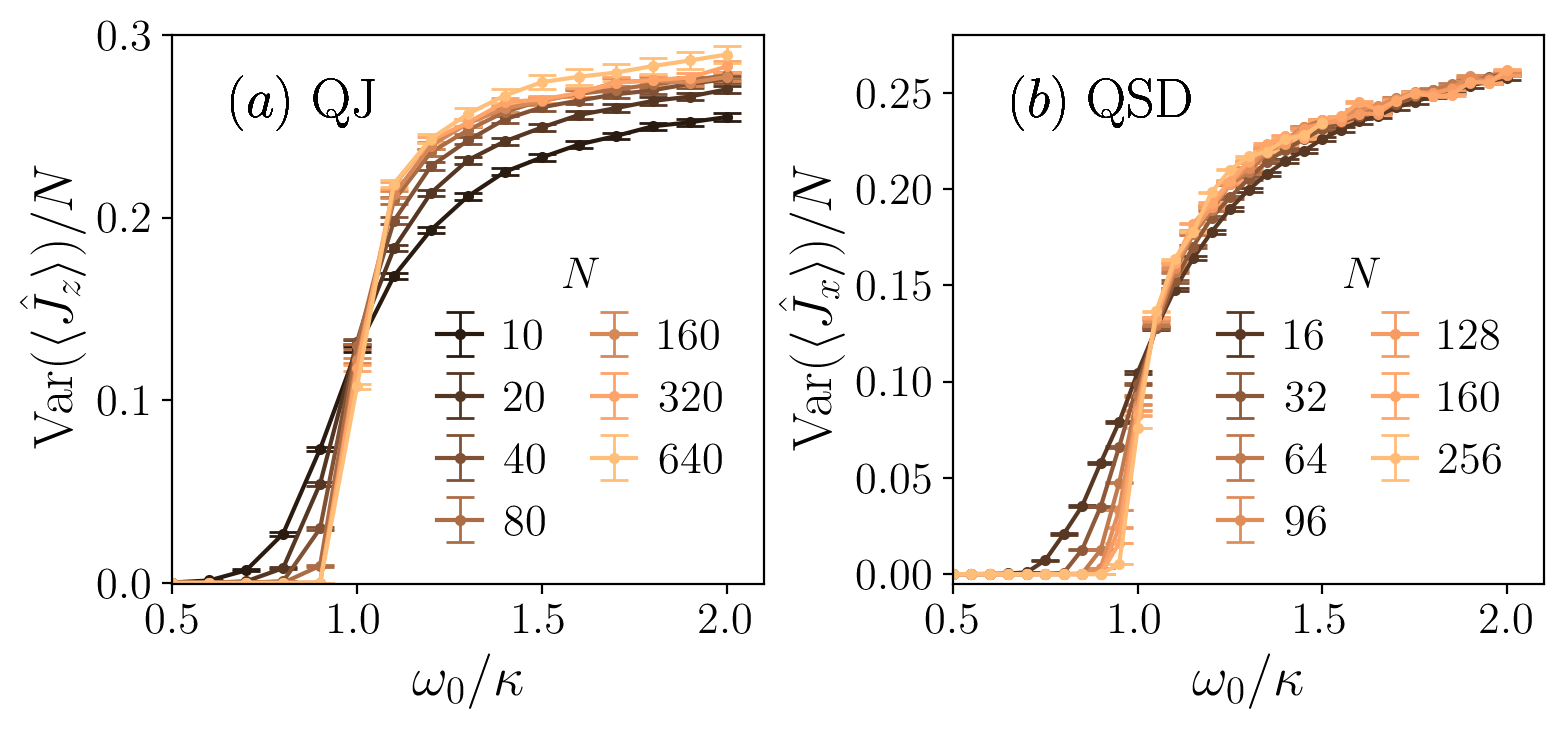}
    \caption{(a) Rescaled trajectory variance of $\langle \hat{J}_z\rangle$ for various system sizes and along the phase diagram for the quantum jump evolution. The trajectory histogram of $\langle \hat{J}_z\rangle$ is experimentally observable, cf.~Ref.~\cite{ferioli2023anonequilibrium}. 
    (b) Rescaled trajectory variance of $\langle \hat{J}_x\rangle$ for the quantum state diffusion. This quantity is directly obtainable from the homodyne current~\cite{wiseman2009quantummeasurementand}.}
    \label{fig:var}
\end{figure}

{\em Experimental implementation ---}  The model under study here has been recently realized in trapped atomic gases coupled to a mode of the free space electromagnetic environment~\cite{ferioli2021laser,ferioli2023anonequilibrium}. In the sub-wavelength regime, a fraction of the atoms share the same diffraction mode and experience a collective decay, as described by the jump operator in Eq.~\eqref{eq:qj}. 
The effective atom number can be tuned in the experiment by changing the geometry of the setup~\cite{ferioli2023anonequilibrium}.  In a single-shot experiment the number of emitted photons in two orthogonal directions is measured, giving access to the quantum jumps and their statistics. 
In particular, by monitoring the intensity of the light emitted in the direction perpendicular to the cloud it is possible to access the population of the atomic excited states, $\langle \hat{J}_z\rangle$, and its statistics. 
The polynomial cost of post-selection in this system suggests that reconstructing the trajectory histogram of this observable could be doable as in experiments with single qubits~\cite{roch2014observation,campagne2016observing}. In Fig.~\ref{fig:var}(a), we compute the variance of the collective spin magnetization, showing a sharp transition at $\omega_0/\kappa=1$.
An even more direct signature of the transition can be obtained by performing homodyne detection and measuring the variance of the homodyne current $\overline{dI_t^2}\sim \overline{\langle \hat{J}_x\rangle^2}$, cf. Eq.~\eqref{eq:dit} . Hence from the first two moments of  $dI_t$ we can reconstruct the variance of $\langle \hat{J}_x\rangle$, cf.~Fig.~\ref{fig:var}(b). \emph{En passant}, we note that these variances constitute a lower bound of the quantum Fisher information~\cite{pezze2018quantummetrologywith}, thus they partially access multipartite entanglement aspects~\footnote{Nevertheless, no direct relationship is in general present between these variances and the entanglement entropy.}. 
Collective dissipative processes such as those at play here can be  engineered in other platforms such as atoms coupled to a cavity mode~\cite{kessler2021observation} or qubits collectively coupled to a microwave resonator~\cite{wang2020controllable} or a waveguide~\cite{liedl2022observation}. Similar phenomenology is expected also in other  dissipative time crystals models, \emph{e.g.}, in~\cite{Zhu_2019}.

{\em Conclusions ---}            
In this Letter, we discussed a class of many-body system where it is possible to follow the dynamics along quantum trajectories without suffering from the post-selection problem. Specifically we discussed a case, that can be observed in existing experimental platforms~\cite{ferioli2023anonequilibrium,ferioliprivate}, and studied 
its entanglement properties. 
\blue{We expect this behavior is generic in semiclassical dynamics. When the latter breaks down, we expect the post-selection overhead to become again exponential}.

\begin{acknowledgements}

{\em Acknowledgments ---} We would like to thank M.~Dalmonte, F.~Iemini, P.~Sierant, S.~Pappalardi, G.~Fux, and Z.~Li for very fruitful conversations and collaborations on related topics. We acknowledge computational resources on the Coll\`ege de France IPH cluster, the CINECA award under the ISCRA initiative (IsB28\_GAMING and IsCb0\_QUJENU), and from MUR, PON “Ricerca e Innovazione 2014-2020”, 
under Grant No. PIR01\_00011 - (I.Bi.S.Co.). This work was supported by the ANR grant ``NonEQuMat'' (ANR-19-CE47-0001) (X.T. and M.S.), by a Google Quantum Research Award (R.F.), by PNRR MUR project PE0000023- NQSTI (P.L., G.P., A.R., and R.F.), by the European Union’s Horizon 2020 research and innovation programme under Grant Agreement No 101017733, by the MUR project CN\_00000013-ICSC (P.L.),  and by the  QuantERA II Programme STAQS project that has received funding from the European Union’s Horizon 2020 research and innovation programme under Grant Agreement No 101017733 (P.L.). This work is co-funded by the European Union (ERC, RAVE, 101053159) (R.F). Views and opinions expressed are however those of the author(s) only and do not necessarily reflect those of the European Union or the European Research Council. Neither the European  Union nor the granting authority can be held responsible for them.

\end{acknowledgements}


%

\clearpage
\appendix

\begin{widetext}
	\begin{center}
		\textbf{\large \centering Supplemental Material: \\ Many-body Dynamics in Monitored Atomic Gases Without Post-Selection Barrier }
	\end{center}
\end{widetext}

\setcounter{equation}{0}
\setcounter{figure}{0}
\setcounter{table}{0}
\setcounter{page}{1}
\renewcommand{\theequation}{S\arabic{equation}}
\setcounter{figure}{0}
\renewcommand{\thefigure}{S\arabic{figure}}
\renewcommand{\thepage}{S\arabic{page}}
\renewcommand{\thesection}{S\arabic{section}}
\renewcommand{\thetable}{S\arabic{table}}
\makeatletter

\renewcommand{\thesection}{\arabic{section}}
\renewcommand{\thesubsection}{\thesection.\arabic{subsection}}
\renewcommand{\thesubsubsection}{\thesubsection.\arabic{subsubsection}}

In this Supplemental Material, we discuss: details on the numerical implementation for the stochastic Schr\"odinger equations considered, additional numerical details, and a phenomenological argument for the saturation time logarithmic in system sizes. {Moreover, we study an interacting model with power-law decaying $ZZ$ interactions and demonstrate that the results of the main text are not fine-tuned since they hold whenever the interactions are sufficiently long-ranged.}

\section{Numerical implementation}
We simulate the quantum jump (QJ) evolution and the quantum state diffusion (QSD) with exact numerical methods that exploit the permutation symmetry of the system. Indeed, the operator $\hat{J}_\alpha$ for $\alpha=x,y,z$ are invariant under exchanges of its constituent. 
Consequently, the wave function complexity is restricted from the full $2^N$-dimensional Hilbert space to the subspace $N+1$ generated by the Dicke states~\cite{shammah2018openquantumsystems}. 
Given the permutationally symmetric pure state $|\Psi\rangle$ (or $\hat{\rho}=|\Psi\rangle\langle \Psi|$), we compute the entanglement entropy as described in Refs.~\cite{latorre2005entanglemententropyin,lerose2020originofthe,lerose2020bridgingentanglementdynamics}. 
The quantum Fisher information is obtained by explicitly evaluating Eq.~(7) on the given density matrix $\hat{\rho}$. 
We now present additional details for the implementation and the choice of hyperparameters considered. 
We start from the fully polarized Dicke state, but we have tested that the stationary state properties are independent of the initial state.

\textit{Quantum jump evolution ---}
For the QJ setup, we simulate the system dynamics in the Dicke basis. In the figures of the main text, averages are computed over $\mathcal{N} = 1000$ trajectories using an integration time step of $\kappa \delta t = 0.002 / N$, which is enough to achieve convergence for all analyzed system sizes. We compute the infinitesimal time-evolution operator of the “no jump” phase,
\begin{equation}\label{eq:evo}
	\hat{U}(\delta t) = e^{-i \hat{\mathcal{H}}_\text{nj} \delta t},
\end{equation}
using the Pad\'e approximation technique implemented in the \textsf{expokit} library~\cite{expokit}, and propagate the state by repeated applications of $\hat{U}(\delta t)$. See the main text for the definition of $\hat{\mathcal{H}}_\text{nj}$. We then use the waiting-time distribution algorithm to determine the occurrence of quantum jumps~\cite{daley}. In particular, the algorithm works as follows: A random number $r$ is extracted from the uniform distribution $\mathcal{U}([0,1])$. The system state is propagated using the time evolution operator of Eq.~\eqref{eq:evo}. Since $\hat{\mathcal{H}}_\text{nj}$ is non-Hermitian, the norm of the time-evolved state decreases in time, resulting in an unnormalized state $\ket{\tilde\psi(t)}$. At $t = t^*$, where $\braket{\tilde\psi(t^*) | \tilde \psi(t^*)} = r$, a quantum jump occurs and the system state becomes Eq.~(2) of the main text. A new random number $r$ is extracted and the algorithm is repeated until the the state is propagated to the final time.

We evolve our systems up to $\kappa T = 400$ and compute the long-time averaged half-chain entanglement entropy as
\begin{equation}\label{eq:lta}
	\overline{S_{N/2}} = \frac{1}{\Delta T} \int_{T - \Delta T}^T S_{N/2}(t) \, dt,
\end{equation}
with $\kappa \Delta T = 100$. 

\begin{figure*}[t]
	\centering
	\includegraphics[width=\textwidth]{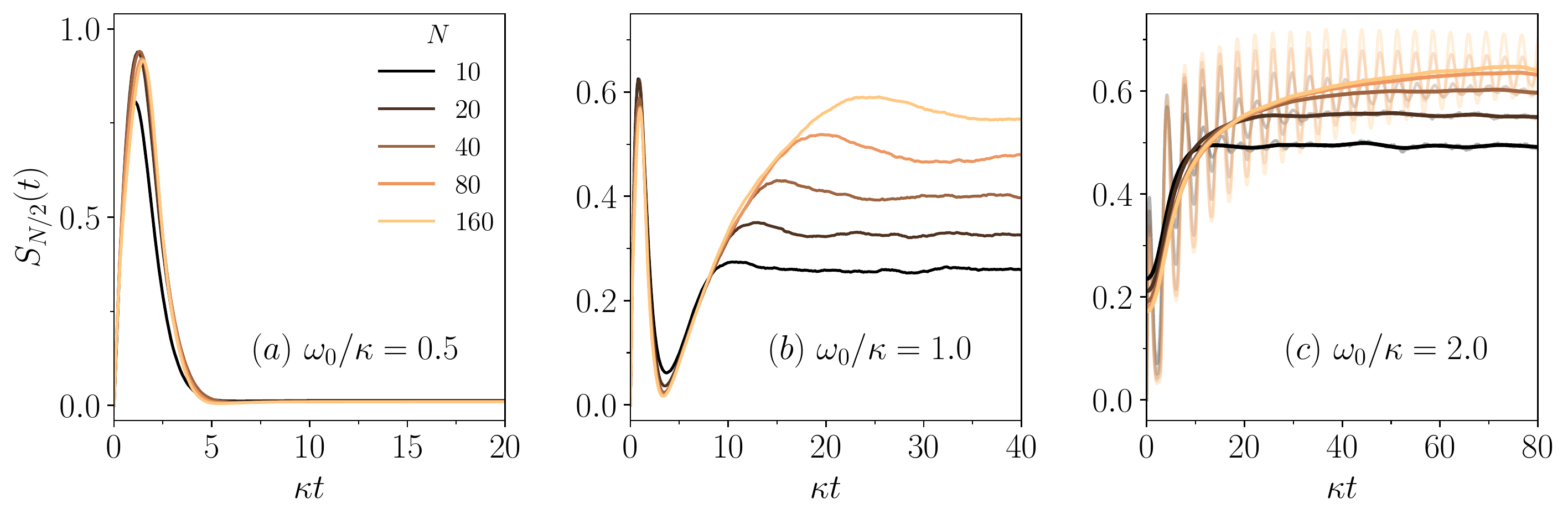}
	\caption{QJ simulations. Dynamics of the half-chain entanglement entropy $S_{N/2}(t)$ for several values of $N$. (a)~$\omega_0  /\kappa = 0.5$. (b)~$\omega_0  /\kappa = 1.0$. (c)~$\omega_0  /\kappa = 2.0$. In (c), transparent lines are real dynamics and solid lines are Gaussian-filtered data with radius $\sigma = 20$.}
	\label{fig:btc-entanglement-dynamics}
\end{figure*}

\textit{Quantum state diffusion ---} 
For the QSD setup, we study the evolution of the density matrix with methods analogous to the quantum jump evolution. At each time step $\delta t$, we exponentiate the dynamical evolution in Eq.~(5) using the Pad\'e approximation technique in the \textsf{expokit} library~\cite{expokit}.
Importantly, because of the \^Ito rules, the dynamical generator is~\cite{wiseman2009quantummeasurementand}
\begin{equation}
	\hat{V}_t = V_0 e^{-i \hat{{\cal H}} \delta t - \frac{ \kappa \delta t}{J}\left(\hat{J}_x\hat{J}_- - 2 \hat{J}_- \langle \hat{J}_x\rangle_t \right) + \sqrt{\frac{\kappa \eta}{J}} dW \hat{J}_-   },
\end{equation} 
with $V_0$ a normalization constant, and the density matrix is given by $\hat{\rho}_{\mathrm{w},t+\delta t}= \hat{V}_t \hat{\rho}_{\mathrm{w},t} \hat{V}^\dagger_t/\mathrm{tr}(\hat{V}_t \hat{\rho}_{\mathrm{w},t} \hat{V}^\dagger_t)$. 
In this case, the complexity is rescaled from $2^N\times 2^N$ to $(N+1)\times (N+1)$. 
For the simulation presented in the main text, we consider $\delta t=5 \times 10^{-3}$ that is sufficient for convergence, as we have tested considering $\delta t=10^{-2}$ and $\delta t=10^{-4}$ giving quantitatively stable results. We consider $\mathcal{N}=1000$ realizations for $N\le 96$, $\mathcal{N}=250$ for $N=128,160$ and $\mathcal{N}=100$ for $N=256$. 
The maximum time evolution is $T = 400/\kappa$, and the average over the trajectories and the variance are obtained over the times $t\ge 300/\kappa$ and all quantum trajectories. 

\section{Additional numerical results}
We complement the numerical findings in the main paper with additional data.

\begin{figure}[h!]
	\centering
	\includegraphics[width=\columnwidth]{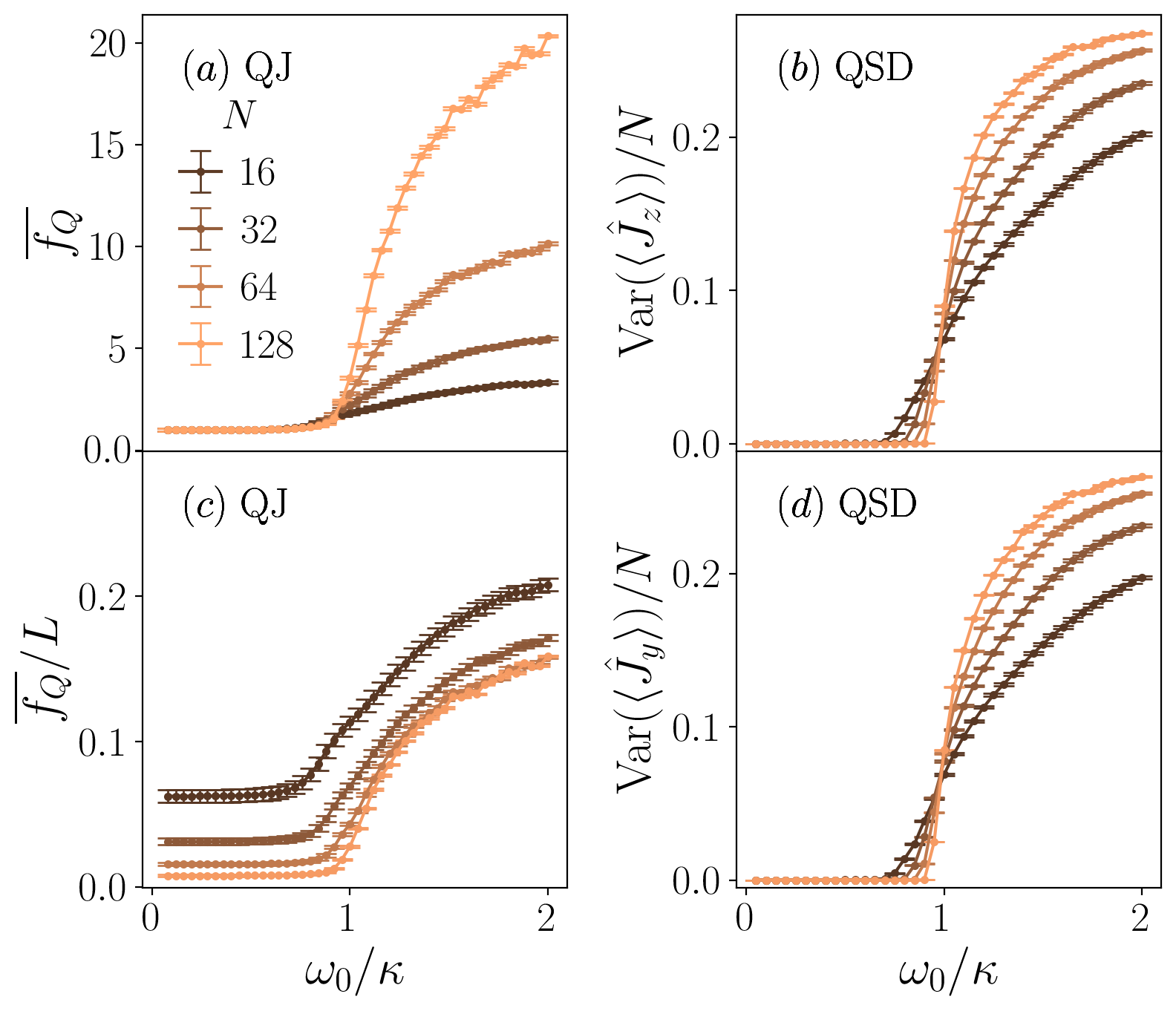}
	\caption{(a)~Average Fisher density for the quantum jump evolution at varying values of $\omega_0/\kappa$ and system sizes $N$. The normal phase and the time crystal phase correspond to $\overline{f_Q}\simeq 1$ and $\overline{f_Q}\simeq N$ respectively. We highlight this scaling by plotting $\overline{f_Q}/N$ in (c). 
		Rescaled variance of the operator $\hat{J}_z$ (b) and $\hat{J}_y$ (d) over the trajectories. Both these quantifier develop a sharp transition at the critical point $\omega_0=\kappa$. }
	\label{fig:supp}
\end{figure}

First, we present other numerical results for the early-time quantum jump evolution of the half-system entanglement entropy averaged over $\mathcal{N} = 10000$ trajectories. These results are presented in Fig.~\ref{fig:btc-entanglement-dynamics}.
Panel~(a) demonstrates that the dynamic is rapidly saturating in the normal phase (exemplified by $\omega_0/\kappa = 0.5$) to a value close to zero. The state indeed is close to an eigenstate of $\hat{J}_-$ with low quantum information encoded. The saturation time is independent on $N$.
Conversely, panel~(c) shows the oscillatory effect that characterizes the time-crystal phase of the system, as presented for $\omega_0/\kappa = 2$. 
These regimes are separated by the transition at $\omega_0=\kappa$, where a non-trivial scaling dynamics is observed, cf.~panel~(b).

In order to compute the saturation time $\tau$, we proceed in two different ways for $\omega_0/\kappa = 1$ and $\omega_0/\kappa > 1$. At the critical point $\omega_0/\kappa = 1$, we evaluate the entanglement entropy at the plateau as in Eq.~\eqref{eq:lta}, we then compute the intersection between this value and the crescent branch of $S_{N/2}(t)$ and use this as an estimate of $\tau$. In the time-crystal phase, we use a Gaussian filter with radius $\sigma = 20$ to filter out the time crystal oscillations and reduce the dynamics to an asymptotic saturation to $\overline{S_{N/2}}$ according to the law $S_{N/2}(t) = a + b \exp (-t/\tau)$, and estimate $\tau$ by fitting the filtered data. The results of our analysis are reported in Fig.~1(d) of the main text.

Furthermore, we study the quantum Fisher information of the quantum jump unraveling in Fig.~\ref{fig:supp}(a). As we see, the QFI density $\overline{f_Q}$ separates a trivial phase ($\overline{f_Q}\simeq 1$) at $\omega_0<\kappa$ from a maximally multipartite phase at $\omega_0>\kappa$. To highlight that $\overline{f_Q}\sim N$ we plot the rescaled $\overline{f_Q}/N$ in Fig.~\ref{fig:supp}(c). Already for $N=64$, this quantity saturates to a limit curve, highlighting the robustness of the phase.
This shows the quantum jump evolution has an inequivalent multipartite entanglement pattern compared to the quantum state diffusion, cf.~main text.

Lastly, we study the variance of $\langle \hat{J}_z\rangle$ and $\langle \hat{J}_y\rangle$ in the quantum state diffusion, see Fig.~\ref{fig:supp}(b) and (d) respectively. 
While these objects are more involved from an experimental perspective compared to $\mathrm{Var}(\langle \hat{J}_x\rangle)$, cf.~main text, they also reveal a sharp transition at the critical point $\omega_0/\kappa$. In particular, the variance over the $z$-direction reveals qualitatively similar features to the QJ case, cf.~Fig.~3 in the main text.

\subsection{Saturation Time}

The logarithmic scaling of the saturation time is a signature of collective dissipation~\cite{gross1982superradiance}. A simple estimate can be obtained by assuming that the coherent drive pushes the system at the top of the Dicke ladder while the collective dissipation drives the system down the ladder from a state $M$ to $M-1$  by emitting a photon through a quantum jump. The typical time scale for this process can be obtained from the imaginary part of the non-Hermitian Hamiltonian $\Gamma_{M,M-1}=(\kappa/2J)\langle J_+J_-\rangle=(\kappa/2J)(J+M)(J-M+1)$. The saturation time is given by summing over the jump events $\tau\sim\sum_{M}1/\Gamma_{M,M-1}$. 
For single particle jumps, independent on the state $M$ of the system, this would give $\Gamma\sim \kappa$ as expected for independent Poisson processes and a saturation time $\tau\sim N$. The key point is that in our case the emission is a correlated process, \textit{i.\,e.}, its rate depends on how many jumps have occurred before through $\Gamma_{M,M-1}$, and emission gets faster as the system goes down in the ladder. A simple calculation gives 
\begin{align}
	\tau &=\frac{2J}{\kappa}\sum_{M=J}^{-J+1}\frac{1}{(J+M)(J-M+1)}=\\
	&=\frac{4J}{\kappa(N+1)}\sum_{M=J}^{-J+1}\left(\frac{1}{J+M}+\frac{1}{J-M+1}\right)=\\
	&=\frac{8J}{\kappa(N+1)}\sum_{s=1}^N \frac{1}{s}
\end{align}
which, in the large $N$ limit, scales as $\tau\sim \ln N$.
We note that the above argument is approximate in that the coherent drive is only used to prepare the state at the top of the ladder, but does not enter the saturation time --- this should work in the strong drive regime $\omega_0\gg \kappa$.

\begin{figure}[tb]
	\centering
	\includegraphics[width=\columnwidth]{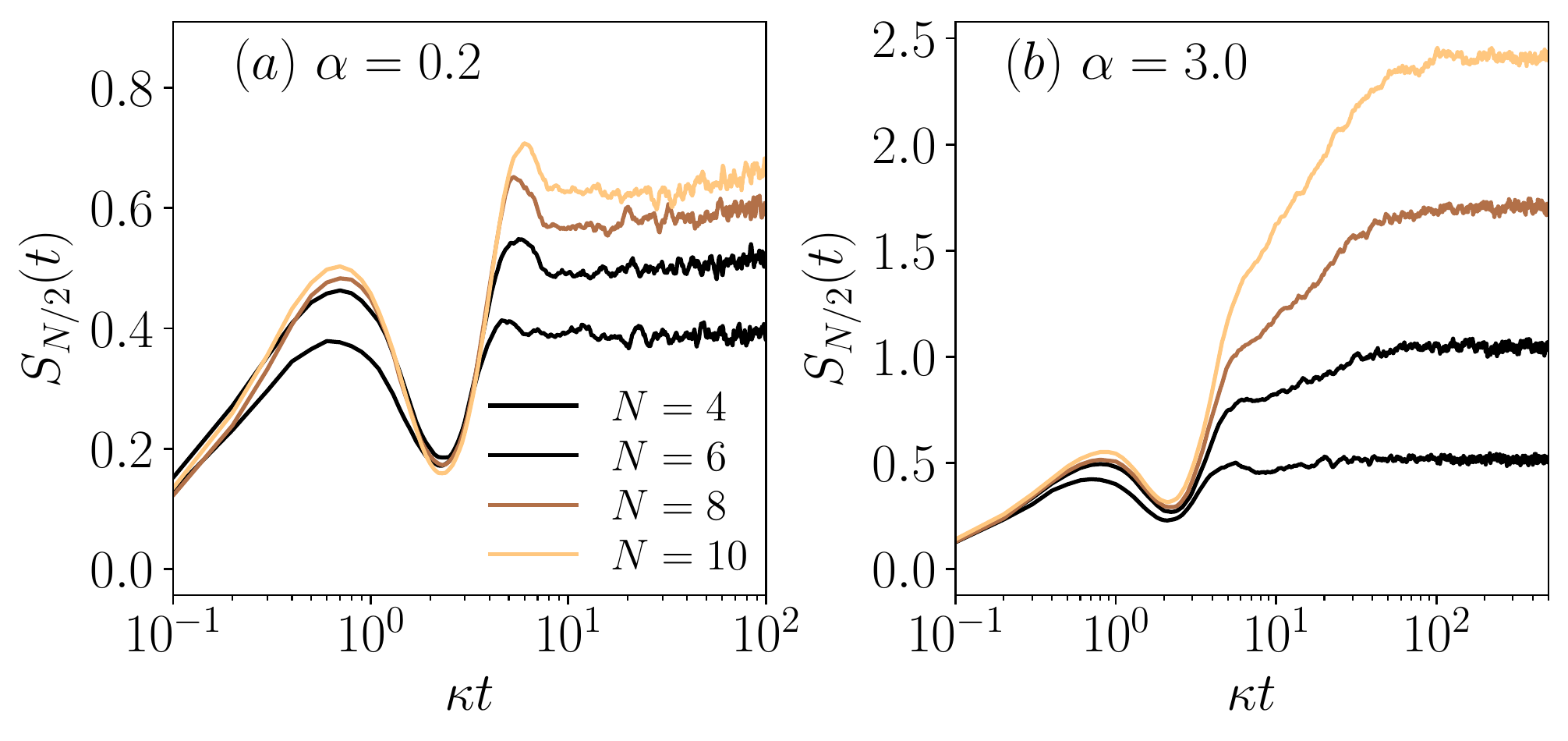}
	\caption{\add{Dynamics of the half-chain entanglement entropy $S_{N/2}(t)$ for the power-law model [Eq.~\eqref{eq:pl}]. (a) $\alpha = 0.2$, long-range regime. (b) $\alpha = 3.0$, short-range regime. Parameters: $\omega_0/\kappa = 1.5$, $\omega_z/\kappa = 0.5$.}}
	\label{fig:ent-pl}
\end{figure}

\subsection{Power-law interacting model}

\add{In this section, we consider a generalization of the model presented in the main text. We use it to demonstrate that the results therein are robust with respect to perturbations of the considered model, provided the resulting Hamiltonian is sufficiently long-range. We consider the following Hamiltonian,
	\begin{equation}
		\label{eq:pl}
		\hat{\mathcal{H}} = \omega_0 \hat{J}_x + \frac{\omega_z}{\mathcal{Z}_{\alpha, N}} \sum_{j > i} \frac{\hat{\sigma}^z_i \hat{\sigma}^z_j}{{d(i, j)}^\alpha},
	\end{equation}
	where $\omega_z$ denotes the interaction strength, $d(i, j) = \min(\lvert i - j \rvert, N - \lvert i - j \rvert)$ is the distance between sites $i$ and $j$ along the chain (with periodic boundary conditions) and the exponent $\alpha$ determines the interaction range. The Kac normalization factor $\mathcal{Z}_{\alpha, N} = N^{-1} \sum_{i \ne j} {d(i, j)}^{-\alpha} $ ensures the Hamiltonian is extensive in the thermodynamic limit for all values of $\alpha$. In particular, the infinite-range limit $\alpha = 0$ is equivalent to the Lipkin-Meshkov-Glick model (considered in the main text), while the opposite limit $\alpha = \infty$ corresponds to the transverse-field Ising model with nearest-neighbor interactions. The $\alpha < 1$ regime is denoted as long-range regime: in the thermodynamic limit, the model behaves as its infinite-range limit $\alpha = 0$. Larger values of $\alpha$ result in a short-range model instead. Thus, this Hamiltonian allows us to tune the interaction range and continuously depart from the infinite-range model considered in the main text. }

\add{On top of this model, we add collective dissipation as done previously and study the dynamics of the half-chain entanglement entropy $S_{N/2}$ following the quantum jump unravelling, for different values of $\alpha$. Due to the fact that the Hamiltonian is no longer $\mathrm{SU}(2)$-symmetric, only small systems are amenable to simulations via exact diagonalization in the full Hilbert space [$N \sim O(10)$], which is the approach we follow here. We focus our attention on $\omega_0 / \kappa = 1.5$, $\omega_z/\kappa = 0.5$, $\alpha \in \left\lbrace 0.2, 2.0 \right\rbrace$. The initial state is the fully polarized Dicke state; results are averaged over $\mathcal{N} = 1000 $ trajectories.} 

\begin{figure}[tb]
	\centering
	\includegraphics[width=\columnwidth]{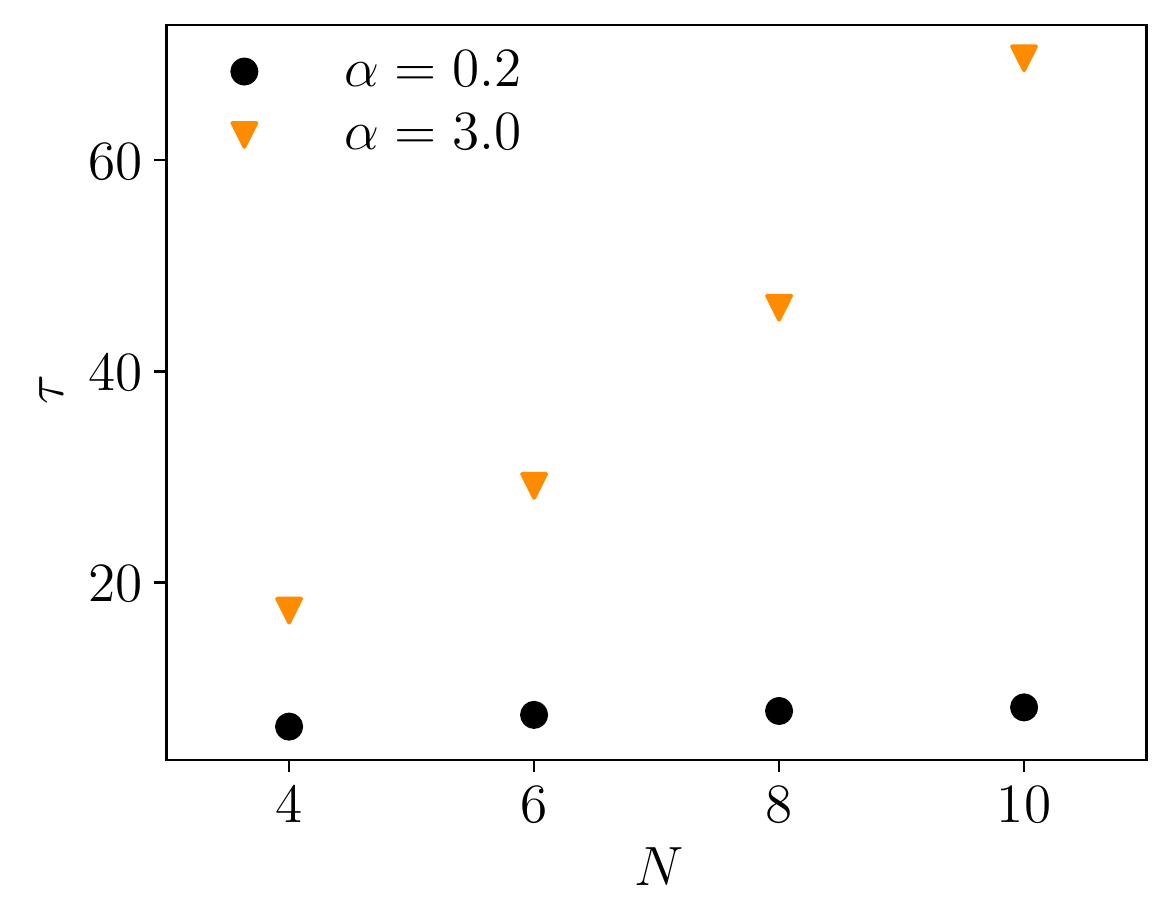}
	\caption{\add{Saturation time $\tau$ versus $N$ in the power-law model [Eq.~\eqref{eq:pl}] for different interaction ranges. Parameters: $\omega_0/\kappa = 1.5$, $\omega_z/\kappa = 0.5$.}}
	\label{fig:pl-tau}
\end{figure}

\add{In Fig.~\ref{fig:ent-pl} we plot $S_{N/2}(t)$ versus $t$ for the two considered values of $\alpha$. In both cases, the entanglement entropy reaches a plateau, persisting up to very long times. However, the important difference to remark between the two cases is that the saturation time $\tau$ at which the plateau is reached scales very differently with the system size. This fact is highlighted in Fig.~\ref{fig:pl-tau}, where the difference between the two regimes is evident already at these very small sizes. Remarkably, in the long-range regime we observe features similar to the infinite-range limit, \textit{i.\,e.}, a very weak dependence of $\tau$ from the system size $N$, which is compatible with the logarithmic scaling observed in the main text for the infinite-range model where much larger system sizes are accessible with numerical simulations. Thus, we conclude that, in the long-range regime, the system remains free from the postselection problem up to very long times. (We note this is a prethermal phase, which disappears at time $t\gg \tau$). By contrast, when $\alpha$ is decreased and the model becomes short-ranged, this conclusion no longer holds as the saturation time scales more rapidly with $N$.}


\begin{thebibliography}{118}%
\makeatletter
\providecommand \@ifxundefined [1]{%
 \@ifx{#1\undefined}
}%
\providecommand \@ifnum [1]{%
 \ifnum #1\expandafter \@firstoftwo
 \else \expandafter \@secondoftwo
 \fi
}%
\providecommand \@ifx [1]{%
 \ifx #1\expandafter \@firstoftwo
 \else \expandafter \@secondoftwo
 \fi
}%
\providecommand \natexlab [1]{#1}%
\providecommand \enquote  [1]{``#1''}%
\providecommand \bibnamefont  [1]{#1}%
\providecommand \bibfnamefont [1]{#1}%
\providecommand \citenamefont [1]{#1}%
\providecommand \href@noop [0]{\@secondoftwo}%
\providecommand \href [0]{\begingroup \@sanitize@url \@href}%
\providecommand \@href[1]{\@@startlink{#1}\@@href}%
\providecommand \@@href[1]{\endgroup#1\@@endlink}%
\providecommand \@sanitize@url [0]{\catcode `\\12\catcode `\$12\catcode
  `\&12\catcode `\#12\catcode `\^12\catcode `\_12\catcode `\%12\relax}%
\providecommand \@@startlink[1]{}%
\providecommand \@@endlink[0]{}%
\providecommand \url  [0]{\begingroup\@sanitize@url \@url }%
\providecommand \@url [1]{\endgroup\@href {#1}{\urlprefix }}%
\providecommand \urlprefix  [0]{URL }%
\providecommand \Eprint [0]{\href }%
\providecommand \doibase [0]{https://doi.org/}%
\providecommand \selectlanguage [0]{\@gobble}%
\providecommand \bibinfo  [0]{\@secondoftwo}%
\providecommand \bibfield  [0]{\@secondoftwo}%
\providecommand \translation [1]{[#1]}%
\providecommand \BibitemOpen [0]{}%
\providecommand \bibitemStop [0]{}%
\providecommand \bibitemNoStop [0]{.\EOS\space}%
\providecommand \EOS [0]{\spacefactor3000\relax}%
\providecommand \BibitemShut  [1]{\csname bibitem#1\endcsname}%
\let\auto@bib@innerbib\@empty
\bibitem [{\citenamefont
  {Carmichael}(1999)}]{carmichael1999statisticalmethodsin}%
  \BibitemOpen
  \bibfield  {author} {\bibinfo {author} {\bibfnamefont {H.}~\bibnamefont
  {Carmichael}},\ }\href@noop {} {\emph {\bibinfo {title} {Statistical Methods
  in Quantum Optics 1}}}\ (\bibinfo  {publisher} {Springer Science \& Business
  Media},\ \bibinfo {year} {Berlin, Germany, 1999})\BibitemShut {NoStop}%
\bibitem [{\citenamefont {Wiseman}\ and\ \citenamefont
  {Milburn}(2009)}]{wiseman2009quantummeasurementand}%
  \BibitemOpen
  \bibfield  {author} {\bibinfo {author} {\bibfnamefont {H.~M.}\ \bibnamefont
  {Wiseman}}\ and\ \bibinfo {author} {\bibfnamefont {G.~J.}\ \bibnamefont
  {Milburn}},\ }\href@noop {} {\emph {\bibinfo {title} {Quantum Measurement and
  Control}}}\ (\bibinfo  {publisher} {Cambridge University Press},\ \bibinfo
  {year} {Cambridge, England, 2009})\BibitemShut {NoStop}%
\bibitem [{\citenamefont {Jacobs}(2014)}]{jacobs2014quantummeasurementtheory}%
  \BibitemOpen
  \bibfield  {author} {\bibinfo {author} {\bibfnamefont {K.}~\bibnamefont
  {Jacobs}},\ }\href@noop {} {\emph {\bibinfo {title} {Quantum Measurement
  Theory and its Applications}}}\ (\bibinfo  {publisher} {Cambridge University
  Press},\ \bibinfo {year} {Cambridge, England, 2014})\BibitemShut {NoStop}%
\bibitem [{\citenamefont {Li}\ \emph {et~al.}(2018)\citenamefont {Li},
  \citenamefont {Chen},\ and\ \citenamefont
  {Fisher}}]{li2018quantumzenoeffect}%
  \BibitemOpen
  \bibfield  {author} {\bibinfo {author} {\bibfnamefont {Y.}~\bibnamefont
  {Li}}, \bibinfo {author} {\bibfnamefont {X.}~\bibnamefont {Chen}},\ and\
  \bibinfo {author} {\bibfnamefont {M.~P.~A.}\ \bibnamefont {Fisher}},\ }\href
  {https://doi.org/10.1103/PhysRevB.98.205136} {\bibfield  {journal} {\bibinfo
  {journal} {Phys. Rev. B}\ }\textbf {\bibinfo {volume} {98}},\ \bibinfo
  {pages} {205136} (\bibinfo {year} {2018})}\BibitemShut {NoStop}%
\bibitem [{\citenamefont {Skinner}\ \emph {et~al.}(2019)\citenamefont
  {Skinner}, \citenamefont {Ruhman},\ and\ \citenamefont
  {Nahum}}]{skinner2019measurementinducedphase}%
  \BibitemOpen
  \bibfield  {author} {\bibinfo {author} {\bibfnamefont {B.}~\bibnamefont
  {Skinner}}, \bibinfo {author} {\bibfnamefont {J.}~\bibnamefont {Ruhman}},\
  and\ \bibinfo {author} {\bibfnamefont {A.}~\bibnamefont {Nahum}},\ }\href
  {https://doi.org/10.1103/PhysRevX.9.031009} {\bibfield  {journal} {\bibinfo
  {journal} {Phys. Rev. X}\ }\textbf {\bibinfo {volume} {9}},\ \bibinfo {pages}
  {031009} (\bibinfo {year} {2019})}\BibitemShut {NoStop}%
\bibitem [{\citenamefont {Fisher}\ \emph {et~al.}(2023)\citenamefont {Fisher},
  \citenamefont {Khemani}, \citenamefont {Nahum},\ and\ \citenamefont
  {Vijay}}]{fisher2023randomquantumcircuits}%
  \BibitemOpen
  \bibfield  {author} {\bibinfo {author} {\bibfnamefont {M.~P.}\ \bibnamefont
  {Fisher}}, \bibinfo {author} {\bibfnamefont {V.}~\bibnamefont {Khemani}},
  \bibinfo {author} {\bibfnamefont {A.}~\bibnamefont {Nahum}},\ and\ \bibinfo
  {author} {\bibfnamefont {S.}~\bibnamefont {Vijay}},\ }\href
  {https://doi.org/10.1146/annurev-conmatphys-031720-030658} {\bibfield
  {journal} {\bibinfo  {journal} {Annu. Rev. Condens. Matter Phys.}\ }\textbf
  {\bibinfo {volume} {14}},\ \bibinfo {pages} {335} (\bibinfo {year}
  {2023})}\BibitemShut {NoStop}%
\bibitem [{\citenamefont {Potter}\ and\ \citenamefont
  {Vasseur}(2022)}]{potter2022quantumsciencesandtechnology}%
  \BibitemOpen
  \bibfield  {author} {\bibinfo {author} {\bibfnamefont {A.~C.}\ \bibnamefont
  {Potter}}\ and\ \bibinfo {author} {\bibfnamefont {R.}~\bibnamefont
  {Vasseur}},\ }\href@noop {} {\emph {\bibinfo {title} {Quantum Sciences and
  Technology}}}\ (\bibinfo  {publisher} {Springer, Cham},\ \bibinfo {year}
  {2022})\ p.\ \bibinfo {pages} {211}\BibitemShut {NoStop}%
\bibitem [{\citenamefont {Lunt}\ \emph {et~al.}(2022)\citenamefont {Lunt},
  \citenamefont {Richter},\ and\ \citenamefont
  {Pal}}]{lunt2022quantumsciencesandtechnology}%
  \BibitemOpen
  \bibfield  {author} {\bibinfo {author} {\bibfnamefont {O.}~\bibnamefont
  {Lunt}}, \bibinfo {author} {\bibfnamefont {J.}~\bibnamefont {Richter}},\ and\
  \bibinfo {author} {\bibfnamefont {A.}~\bibnamefont {Pal}},\ }\href@noop {}
  {\emph {\bibinfo {title} {Quantum Sciences and Technology}}}\ (\bibinfo
  {publisher} {Springer, Cham},\ \bibinfo {year} {2022})\ p.\ \bibinfo {pages}
  {251}\BibitemShut {NoStop}%
\bibitem [{\citenamefont {Li}\ \emph {et~al.}(2019)\citenamefont {Li},
  \citenamefont {Chen},\ and\ \citenamefont
  {Fisher}}]{li2019measurementdrivenentanglement}%
  \BibitemOpen
  \bibfield  {author} {\bibinfo {author} {\bibfnamefont {Y.}~\bibnamefont
  {Li}}, \bibinfo {author} {\bibfnamefont {X.}~\bibnamefont {Chen}},\ and\
  \bibinfo {author} {\bibfnamefont {M.~P.~A.}\ \bibnamefont {Fisher}},\ }\href
  {https://doi.org/10.1103/PhysRevB.100.134306} {\bibfield  {journal} {\bibinfo
   {journal} {Phys. Rev. B}\ }\textbf {\bibinfo {volume} {100}},\ \bibinfo
  {pages} {134306} (\bibinfo {year} {2019})}\BibitemShut {NoStop}%
\bibitem [{\citenamefont {Szyniszewski}\ \emph {et~al.}(2019)\citenamefont
  {Szyniszewski}, \citenamefont {Romito},\ and\ \citenamefont
  {Schomerus}}]{szyniszewski2019entanglementtransitionfrom}%
  \BibitemOpen
  \bibfield  {author} {\bibinfo {author} {\bibfnamefont {M.}~\bibnamefont
  {Szyniszewski}}, \bibinfo {author} {\bibfnamefont {A.}~\bibnamefont
  {Romito}},\ and\ \bibinfo {author} {\bibfnamefont {H.}~\bibnamefont
  {Schomerus}},\ }\href {https://doi.org/10.1103/PhysRevB.100.064204}
  {\bibfield  {journal} {\bibinfo  {journal} {Phys. Rev. B}\ }\textbf {\bibinfo
  {volume} {100}},\ \bibinfo {pages} {064204} (\bibinfo {year}
  {2019})}\BibitemShut {NoStop}%
\bibitem [{\citenamefont {Jian}\ \emph {et~al.}(2020)\citenamefont {Jian},
  \citenamefont {You}, \citenamefont {Vasseur},\ and\ \citenamefont
  {Ludwig}}]{jian2020measurementinducedcriticality}%
  \BibitemOpen
  \bibfield  {author} {\bibinfo {author} {\bibfnamefont {C.-M.}\ \bibnamefont
  {Jian}}, \bibinfo {author} {\bibfnamefont {Y.-Z.}\ \bibnamefont {You}},
  \bibinfo {author} {\bibfnamefont {R.}~\bibnamefont {Vasseur}},\ and\ \bibinfo
  {author} {\bibfnamefont {A.~W.~W.}\ \bibnamefont {Ludwig}},\ }\href
  {https://doi.org/10.1103/PhysRevB.101.104302} {\bibfield  {journal} {\bibinfo
   {journal} {Phys. Rev. B}\ }\textbf {\bibinfo {volume} {101}},\ \bibinfo
  {pages} {104302} (\bibinfo {year} {2020})}\BibitemShut {NoStop}%
\bibitem [{\citenamefont {{Y. Li, R. Vasseur, M. P. A. Fisher, and A. W. W.
  Ludwig}}()}]{li2021statisticalmechanicsmodel}%
  \BibitemOpen
  \bibfield  {author} {\bibinfo {author} {\bibnamefont {{Y. Li, R. Vasseur, M.
  P. A. Fisher, and A. W. W. Ludwig}}},\ }\href@noop {} {}\Eprint
  {https://arxiv.org/abs/2110.02988} {arXiv:2110.02988} \BibitemShut {NoStop}%
\bibitem [{\citenamefont {Zabalo}\ \emph {et~al.}(2020)\citenamefont {Zabalo},
  \citenamefont {Gullans}, \citenamefont {Wilson}, \citenamefont
  {Gopalakrishnan}, \citenamefont {Huse},\ and\ \citenamefont
  {Pixley}}]{zabalo202criticalpropertiesof}%
  \BibitemOpen
  \bibfield  {author} {\bibinfo {author} {\bibfnamefont {A.}~\bibnamefont
  {Zabalo}}, \bibinfo {author} {\bibfnamefont {M.~J.}\ \bibnamefont {Gullans}},
  \bibinfo {author} {\bibfnamefont {J.~H.}\ \bibnamefont {Wilson}}, \bibinfo
  {author} {\bibfnamefont {S.}~\bibnamefont {Gopalakrishnan}}, \bibinfo
  {author} {\bibfnamefont {D.~A.}\ \bibnamefont {Huse}},\ and\ \bibinfo
  {author} {\bibfnamefont {J.~H.}\ \bibnamefont {Pixley}},\ }\href
  {https://doi.org/10.1103/PhysRevB.101.060301} {\bibfield  {journal} {\bibinfo
   {journal} {Phys. Rev. B}\ }\textbf {\bibinfo {volume} {101}},\ \bibinfo
  {pages} {060301} (\bibinfo {year} {2020})}\BibitemShut {NoStop}%
\bibitem [{\citenamefont {Szyniszewski}\ \emph {et~al.}(2020)\citenamefont
  {Szyniszewski}, \citenamefont {Romito},\ and\ \citenamefont
  {Schomerus}}]{szyniszewski2020universalityofentanglement}%
  \BibitemOpen
  \bibfield  {author} {\bibinfo {author} {\bibfnamefont {M.}~\bibnamefont
  {Szyniszewski}}, \bibinfo {author} {\bibfnamefont {A.}~\bibnamefont
  {Romito}},\ and\ \bibinfo {author} {\bibfnamefont {H.}~\bibnamefont
  {Schomerus}},\ }\href {https://doi.org/10.1103/PhysRevLett.125.210602}
  {\bibfield  {journal} {\bibinfo  {journal} {Phys. Rev. Lett.}\ }\textbf
  {\bibinfo {volume} {125}},\ \bibinfo {pages} {210602} (\bibinfo {year}
  {2020})}\BibitemShut {NoStop}%
\bibitem [{\citenamefont {Turkeshi}\ \emph {et~al.}(2020)\citenamefont
  {Turkeshi}, \citenamefont {Fazio},\ and\ \citenamefont
  {Dalmonte}}]{turkeshi2020measurementinducedcriticality}%
  \BibitemOpen
  \bibfield  {author} {\bibinfo {author} {\bibfnamefont {X.}~\bibnamefont
  {Turkeshi}}, \bibinfo {author} {\bibfnamefont {R.}~\bibnamefont {Fazio}},\
  and\ \bibinfo {author} {\bibfnamefont {M.}~\bibnamefont {Dalmonte}},\ }\href
  {https://doi.org/10.1103/PhysRevB.102.014315} {\bibfield  {journal} {\bibinfo
   {journal} {Phys. Rev. B}\ }\textbf {\bibinfo {volume} {102}},\ \bibinfo
  {pages} {014315} (\bibinfo {year} {2020})}\BibitemShut {NoStop}%
\bibitem [{\citenamefont {Lunt}\ \emph {et~al.}(2021)\citenamefont {Lunt},
  \citenamefont {Szyniszewski},\ and\ \citenamefont
  {Pal}}]{lunt2021measurementinducedcriticality}%
  \BibitemOpen
  \bibfield  {author} {\bibinfo {author} {\bibfnamefont {O.}~\bibnamefont
  {Lunt}}, \bibinfo {author} {\bibfnamefont {M.}~\bibnamefont {Szyniszewski}},\
  and\ \bibinfo {author} {\bibfnamefont {A.}~\bibnamefont {Pal}},\ }\href
  {https://doi.org/10.1103/PhysRevB.104.155111} {\bibfield  {journal} {\bibinfo
   {journal} {Phys. Rev. B}\ }\textbf {\bibinfo {volume} {104}},\ \bibinfo
  {pages} {155111} (\bibinfo {year} {2021})}\BibitemShut {NoStop}%
\bibitem [{\citenamefont {Sierant}\ \emph
  {et~al.}(2022{\natexlab{a}})\citenamefont {Sierant}, \citenamefont
  {Schir\`o}, \citenamefont {Lewenstein},\ and\ \citenamefont
  {Turkeshi}}]{sierant2022measurementinducedphase}%
  \BibitemOpen
  \bibfield  {author} {\bibinfo {author} {\bibfnamefont {P.}~\bibnamefont
  {Sierant}}, \bibinfo {author} {\bibfnamefont {M.}~\bibnamefont {Schir\`o}},
  \bibinfo {author} {\bibfnamefont {M.}~\bibnamefont {Lewenstein}},\ and\
  \bibinfo {author} {\bibfnamefont {X.}~\bibnamefont {Turkeshi}},\ }\href
  {https://doi.org/10.1103/PhysRevB.106.214316} {\bibfield  {journal} {\bibinfo
   {journal} {Phys. Rev. B}\ }\textbf {\bibinfo {volume} {106}},\ \bibinfo
  {pages} {214316} (\bibinfo {year} {2022}{\natexlab{a}})}\BibitemShut
  {NoStop}%
\bibitem [{\citenamefont {Nahum}\ \emph {et~al.}(2021)\citenamefont {Nahum},
  \citenamefont {Roy}, \citenamefont {Skinner},\ and\ \citenamefont
  {Ruhman}}]{nahum2021measurementandentanglement}%
  \BibitemOpen
  \bibfield  {author} {\bibinfo {author} {\bibfnamefont {A.}~\bibnamefont
  {Nahum}}, \bibinfo {author} {\bibfnamefont {S.}~\bibnamefont {Roy}}, \bibinfo
  {author} {\bibfnamefont {B.}~\bibnamefont {Skinner}},\ and\ \bibinfo {author}
  {\bibfnamefont {J.}~\bibnamefont {Ruhman}},\ }\href
  {https://doi.org/10.1103/PRXQuantum.2.010352} {\bibfield  {journal} {\bibinfo
   {journal} {PRX Quantum}\ }\textbf {\bibinfo {volume} {2}},\ \bibinfo {pages}
  {010352} (\bibinfo {year} {2021})}\BibitemShut {NoStop}%
\bibitem [{\citenamefont {Zabalo}\ \emph {et~al.}(2022)\citenamefont {Zabalo},
  \citenamefont {Gullans}, \citenamefont {Wilson}, \citenamefont {Vasseur},
  \citenamefont {Ludwig}, \citenamefont {Gopalakrishnan}, \citenamefont
  {Huse},\ and\ \citenamefont {Pixley}}]{zabalo2022operatorscalingdimensions}%
  \BibitemOpen
  \bibfield  {author} {\bibinfo {author} {\bibfnamefont {A.}~\bibnamefont
  {Zabalo}}, \bibinfo {author} {\bibfnamefont {M.~J.}\ \bibnamefont {Gullans}},
  \bibinfo {author} {\bibfnamefont {J.~H.}\ \bibnamefont {Wilson}}, \bibinfo
  {author} {\bibfnamefont {R.}~\bibnamefont {Vasseur}}, \bibinfo {author}
  {\bibfnamefont {A.~W.~W.}\ \bibnamefont {Ludwig}}, \bibinfo {author}
  {\bibfnamefont {S.}~\bibnamefont {Gopalakrishnan}}, \bibinfo {author}
  {\bibfnamefont {D.~A.}\ \bibnamefont {Huse}},\ and\ \bibinfo {author}
  {\bibfnamefont {J.~H.}\ \bibnamefont {Pixley}},\ }\href
  {https://doi.org/10.1103/PhysRevLett.128.050602} {\bibfield  {journal}
  {\bibinfo  {journal} {Phys. Rev. Lett.}\ }\textbf {\bibinfo {volume} {128}},\
  \bibinfo {pages} {050602} (\bibinfo {year} {2022})}\BibitemShut {NoStop}%
\bibitem [{\citenamefont {Sierant}\ and\ \citenamefont
  {Turkeshi}(2022)}]{sierant2022universalbehaviorbeyond}%
  \BibitemOpen
  \bibfield  {author} {\bibinfo {author} {\bibfnamefont {P.}~\bibnamefont
  {Sierant}}\ and\ \bibinfo {author} {\bibfnamefont {X.}~\bibnamefont
  {Turkeshi}},\ }\href {https://doi.org/10.1103/PhysRevLett.128.130605}
  {\bibfield  {journal} {\bibinfo  {journal} {Phys. Rev. Lett.}\ }\textbf
  {\bibinfo {volume} {128}},\ \bibinfo {pages} {130605} (\bibinfo {year}
  {2022})}\BibitemShut {NoStop}%
\bibitem [{\citenamefont {Chiriac\`o}\ \emph {et~al.}(2023)\citenamefont
  {Chiriac\`o}, \citenamefont {Tsitsishvili}, \citenamefont {Poletti},
  \citenamefont {Fazio},\ and\ \citenamefont
  {Dalmonte}}]{chiriaco2023diagrammaticmethodfor}%
  \BibitemOpen
  \bibfield  {author} {\bibinfo {author} {\bibfnamefont {G.}~\bibnamefont
  {Chiriac\`o}}, \bibinfo {author} {\bibfnamefont {M.}~\bibnamefont
  {Tsitsishvili}}, \bibinfo {author} {\bibfnamefont {D.}~\bibnamefont
  {Poletti}}, \bibinfo {author} {\bibfnamefont {R.}~\bibnamefont {Fazio}},\
  and\ \bibinfo {author} {\bibfnamefont {M.}~\bibnamefont {Dalmonte}},\ }\href
  {https://doi.org/10.1103/PhysRevB.108.075151} {\bibfield  {journal} {\bibinfo
   {journal} {Phys. Rev. B}\ }\textbf {\bibinfo {volume} {108}},\ \bibinfo
  {pages} {075151} (\bibinfo {year} {2023})}\BibitemShut {NoStop}%
\bibitem [{\citenamefont {Klocke}\ and\ \citenamefont
  {Buchhold}(2023)}]{klocke2023majorana}%
  \BibitemOpen
  \bibfield  {author} {\bibinfo {author} {\bibfnamefont {K.}~\bibnamefont
  {Klocke}}\ and\ \bibinfo {author} {\bibfnamefont {M.}~\bibnamefont
  {Buchhold}},\ }\href {https://doi.org/10.1103/PhysRevX.13.041028} {\bibfield
  {journal} {\bibinfo  {journal} {Phys. Rev. X}\ }\textbf {\bibinfo {volume}
  {13}},\ \bibinfo {pages} {041028} (\bibinfo {year} {2023})}\BibitemShut
  {NoStop}%
\bibitem [{\citenamefont {Cao}\ \emph {et~al.}(2019)\citenamefont {Cao},
  \citenamefont {Tilloy},\ and\ \citenamefont {{De
  Luca}}}]{cao2019entanglementina}%
  \BibitemOpen
  \bibfield  {author} {\bibinfo {author} {\bibfnamefont {X.}~\bibnamefont
  {Cao}}, \bibinfo {author} {\bibfnamefont {A.}~\bibnamefont {Tilloy}},\ and\
  \bibinfo {author} {\bibfnamefont {A.}~\bibnamefont {{De Luca}}},\ }\href
  {https://doi.org/10.21468/SciPostPhys.7.2.024} {\bibfield  {journal}
  {\bibinfo  {journal} {SciPost Phys.}\ }\textbf {\bibinfo {volume} {7}},\
  \bibinfo {pages} {024} (\bibinfo {year} {2019})}\BibitemShut {NoStop}%
\bibitem [{\citenamefont {Nahum}\ and\ \citenamefont
  {Skinner}(2020)}]{nahum2020entanglementanddynamics}%
  \BibitemOpen
  \bibfield  {author} {\bibinfo {author} {\bibfnamefont {A.}~\bibnamefont
  {Nahum}}\ and\ \bibinfo {author} {\bibfnamefont {B.}~\bibnamefont
  {Skinner}},\ }\href {https://doi.org/10.1103/PhysRevResearch.2.023288}
  {\bibfield  {journal} {\bibinfo  {journal} {Phys. Rev. Res.}\ }\textbf
  {\bibinfo {volume} {2}},\ \bibinfo {pages} {023288} (\bibinfo {year}
  {2020})}\BibitemShut {NoStop}%
\bibitem [{\citenamefont {Buchhold}\ \emph {et~al.}(2021)\citenamefont
  {Buchhold}, \citenamefont {Minoguchi}, \citenamefont {Altland},\ and\
  \citenamefont {Diehl}}]{buchhold2021effectivetheoryfor}%
  \BibitemOpen
  \bibfield  {author} {\bibinfo {author} {\bibfnamefont {M.}~\bibnamefont
  {Buchhold}}, \bibinfo {author} {\bibfnamefont {Y.}~\bibnamefont {Minoguchi}},
  \bibinfo {author} {\bibfnamefont {A.}~\bibnamefont {Altland}},\ and\ \bibinfo
  {author} {\bibfnamefont {S.}~\bibnamefont {Diehl}},\ }\href
  {https://doi.org/10.1103/PhysRevX.11.041004} {\bibfield  {journal} {\bibinfo
  {journal} {Phys. Rev. X}\ }\textbf {\bibinfo {volume} {11}},\ \bibinfo
  {pages} {041004} (\bibinfo {year} {2021})}\BibitemShut {NoStop}%
\bibitem [{\citenamefont {Jian}\ \emph {et~al.}(2022)\citenamefont {Jian},
  \citenamefont {Bauer}, \citenamefont {Keselman},\ and\ \citenamefont
  {Ludwig}}]{chaoming2022criticalityandentanglement}%
  \BibitemOpen
  \bibfield  {author} {\bibinfo {author} {\bibfnamefont {C.-M.}\ \bibnamefont
  {Jian}}, \bibinfo {author} {\bibfnamefont {B.}~\bibnamefont {Bauer}},
  \bibinfo {author} {\bibfnamefont {A.}~\bibnamefont {Keselman}},\ and\
  \bibinfo {author} {\bibfnamefont {A.~W.~W.}\ \bibnamefont {Ludwig}},\ }\href
  {https://doi.org/10.1103/PhysRevB.106.134206} {\bibfield  {journal} {\bibinfo
   {journal} {Phys. Rev. B}\ }\textbf {\bibinfo {volume} {106}},\ \bibinfo
  {pages} {134206} (\bibinfo {year} {2022})}\BibitemShut {NoStop}%
\bibitem [{\citenamefont {Coppola}\ \emph {et~al.}(2022)\citenamefont
  {Coppola}, \citenamefont {Tirrito}, \citenamefont {Karevski},\ and\
  \citenamefont {Collura}}]{coppola2022growthofentanglement}%
  \BibitemOpen
  \bibfield  {author} {\bibinfo {author} {\bibfnamefont {M.}~\bibnamefont
  {Coppola}}, \bibinfo {author} {\bibfnamefont {E.}~\bibnamefont {Tirrito}},
  \bibinfo {author} {\bibfnamefont {D.}~\bibnamefont {Karevski}},\ and\
  \bibinfo {author} {\bibfnamefont {M.}~\bibnamefont {Collura}},\ }\href
  {https://doi.org/10.1103/PhysRevB.105.094303} {\bibfield  {journal} {\bibinfo
   {journal} {Phys. Rev. B}\ }\textbf {\bibinfo {volume} {105}},\ \bibinfo
  {pages} {094303} (\bibinfo {year} {2022})}\BibitemShut {NoStop}%
\bibitem [{\citenamefont {Fava}\ \emph {et~al.}(2023)\citenamefont {Fava},
  \citenamefont {Piroli}, \citenamefont {Swann}, \citenamefont {Bernard},\ and\
  \citenamefont {Nahum}}]{fava2023nonlinearsigmamodels}%
  \BibitemOpen
  \bibfield  {author} {\bibinfo {author} {\bibfnamefont {M.}~\bibnamefont
  {Fava}}, \bibinfo {author} {\bibfnamefont {L.}~\bibnamefont {Piroli}},
  \bibinfo {author} {\bibfnamefont {T.}~\bibnamefont {Swann}}, \bibinfo
  {author} {\bibfnamefont {D.}~\bibnamefont {Bernard}},\ and\ \bibinfo {author}
  {\bibfnamefont {A.}~\bibnamefont {Nahum}},\ }\href
  {https://doi.org/10.1103/PhysRevX.13.041045} {\bibfield  {journal} {\bibinfo
  {journal} {Phys. Rev. X}\ }\textbf {\bibinfo {volume} {13}},\ \bibinfo
  {pages} {041045} (\bibinfo {year} {2023})}\BibitemShut {NoStop}%
\bibitem [{\citenamefont {Poboiko}\ \emph {et~al.}(2023)\citenamefont
  {Poboiko}, \citenamefont {P\"opperl}, \citenamefont {Gornyi},\ and\
  \citenamefont {Mirlin}}]{poboiko2023theoryoffree}%
  \BibitemOpen
  \bibfield  {author} {\bibinfo {author} {\bibfnamefont {I.}~\bibnamefont
  {Poboiko}}, \bibinfo {author} {\bibfnamefont {P.}~\bibnamefont {P\"opperl}},
  \bibinfo {author} {\bibfnamefont {I.~V.}\ \bibnamefont {Gornyi}},\ and\
  \bibinfo {author} {\bibfnamefont {A.~D.}\ \bibnamefont {Mirlin}},\ }\href
  {https://doi.org/10.1103/PhysRevX.13.041046} {\bibfield  {journal} {\bibinfo
  {journal} {Phys. Rev. X}\ }\textbf {\bibinfo {volume} {13}},\ \bibinfo
  {pages} {041046} (\bibinfo {year} {2023})}\BibitemShut {NoStop}%
\bibitem [{\citenamefont {Jian}\ \emph {et~al.}()\citenamefont {Jian},
  \citenamefont {Shapourian}, \citenamefont {Bauer},\ and\ \citenamefont
  {Ludwig}}]{jian2023measurementinducedentanglement}%
  \BibitemOpen
  \bibfield  {author} {\bibinfo {author} {\bibfnamefont {C.-M.}\ \bibnamefont
  {Jian}}, \bibinfo {author} {\bibfnamefont {H.}~\bibnamefont {Shapourian}},
  \bibinfo {author} {\bibfnamefont {B.}~\bibnamefont {Bauer}},\ and\ \bibinfo
  {author} {\bibfnamefont {A.~W.~W.}\ \bibnamefont {Ludwig}},\ }\href@noop {}
  {}\Eprint {https://arxiv.org/abs/2302.09094} {arXiv:2302.09094} \BibitemShut
  {NoStop}%
\bibitem [{\citenamefont {Merritt}\ and\ \citenamefont
  {Fidkowski}(2023)}]{merritt2023entanglementtransitionswith}%
  \BibitemOpen
  \bibfield  {author} {\bibinfo {author} {\bibfnamefont {J.}~\bibnamefont
  {Merritt}}\ and\ \bibinfo {author} {\bibfnamefont {L.}~\bibnamefont
  {Fidkowski}},\ }\href {https://doi.org/10.1103/PhysRevB.107.064303}
  {\bibfield  {journal} {\bibinfo  {journal} {Phys. Rev. B}\ }\textbf {\bibinfo
  {volume} {107}},\ \bibinfo {pages} {064303} (\bibinfo {year}
  {2023})}\BibitemShut {NoStop}%
\bibitem [{\citenamefont {Alberton}\ \emph {et~al.}(2021)\citenamefont
  {Alberton}, \citenamefont {Buchhold},\ and\ \citenamefont
  {Diehl}}]{alberton2021entanglementtransitionin}%
  \BibitemOpen
  \bibfield  {author} {\bibinfo {author} {\bibfnamefont {O.}~\bibnamefont
  {Alberton}}, \bibinfo {author} {\bibfnamefont {M.}~\bibnamefont {Buchhold}},\
  and\ \bibinfo {author} {\bibfnamefont {S.}~\bibnamefont {Diehl}},\ }\href
  {https://doi.org/10.1103/PhysRevLett.126.170602} {\bibfield  {journal}
  {\bibinfo  {journal} {Phys. Rev. Lett.}\ }\textbf {\bibinfo {volume} {126}},\
  \bibinfo {pages} {170602} (\bibinfo {year} {2021})}\BibitemShut {NoStop}%
\bibitem [{\citenamefont {Turkeshi}\ \emph {et~al.}(2021)\citenamefont
  {Turkeshi}, \citenamefont {Biella}, \citenamefont {Fazio}, \citenamefont
  {Dalmonte},\ and\ \citenamefont
  {Schir\'o}}]{turkeshi2021measurementinducedentanglement}%
  \BibitemOpen
  \bibfield  {author} {\bibinfo {author} {\bibfnamefont {X.}~\bibnamefont
  {Turkeshi}}, \bibinfo {author} {\bibfnamefont {A.}~\bibnamefont {Biella}},
  \bibinfo {author} {\bibfnamefont {R.}~\bibnamefont {Fazio}}, \bibinfo
  {author} {\bibfnamefont {M.}~\bibnamefont {Dalmonte}},\ and\ \bibinfo
  {author} {\bibfnamefont {M.}~\bibnamefont {Schir\'o}},\ }\href
  {https://doi.org/10.1103/PhysRevB.103.224210} {\bibfield  {journal} {\bibinfo
   {journal} {Phys. Rev. B}\ }\textbf {\bibinfo {volume} {103}},\ \bibinfo
  {pages} {224210} (\bibinfo {year} {2021})}\BibitemShut {NoStop}%
\bibitem [{\citenamefont {Turkeshi}\ \emph
  {et~al.}(2022{\natexlab{a}})\citenamefont {Turkeshi}, \citenamefont
  {Dalmonte}, \citenamefont {Fazio},\ and\ \citenamefont
  {Schir{\`o}}}]{turkeshi2022entanglementtransitionsfrom}%
  \BibitemOpen
  \bibfield  {author} {\bibinfo {author} {\bibfnamefont {X.}~\bibnamefont
  {Turkeshi}}, \bibinfo {author} {\bibfnamefont {M.}~\bibnamefont {Dalmonte}},
  \bibinfo {author} {\bibfnamefont {R.}~\bibnamefont {Fazio}},\ and\ \bibinfo
  {author} {\bibfnamefont {M.}~\bibnamefont {Schir{\`o}}},\ }\href
  {https://doi.org/10.1103/PhysRevB.105.L241114} {\bibfield  {journal}
  {\bibinfo  {journal} {Physical Review B}\ }\textbf {\bibinfo {volume}
  {105}},\ \bibinfo {pages} {L241114} (\bibinfo {year}
  {2022}{\natexlab{a}})}\BibitemShut {NoStop}%
\bibitem [{\citenamefont {Piccitto}\ \emph {et~al.}(2022)\citenamefont
  {Piccitto}, \citenamefont {Russomanno},\ and\ \citenamefont
  {Rossini}}]{piccitto2022entanglementtransitionsin}%
  \BibitemOpen
  \bibfield  {author} {\bibinfo {author} {\bibfnamefont {G.}~\bibnamefont
  {Piccitto}}, \bibinfo {author} {\bibfnamefont {A.}~\bibnamefont
  {Russomanno}},\ and\ \bibinfo {author} {\bibfnamefont {D.}~\bibnamefont
  {Rossini}},\ }\href {https://doi.org/10.1103/PhysRevB.105.064305} {\bibfield
  {journal} {\bibinfo  {journal} {Phys. Rev. B}\ }\textbf {\bibinfo {volume}
  {105}},\ \bibinfo {pages} {064305} (\bibinfo {year} {2022})}\BibitemShut
  {NoStop}%
\bibitem [{\citenamefont {Piccitto}\ \emph {et~al.}(2023)\citenamefont
  {Piccitto}, \citenamefont {Russomanno},\ and\ \citenamefont
  {Rossini}}]{piccitto2023entanglementdynamicswith}%
  \BibitemOpen
  \bibfield  {author} {\bibinfo {author} {\bibfnamefont {G.}~\bibnamefont
  {Piccitto}}, \bibinfo {author} {\bibfnamefont {A.}~\bibnamefont
  {Russomanno}},\ and\ \bibinfo {author} {\bibfnamefont {D.}~\bibnamefont
  {Rossini}},\ }\href {https://doi.org/10.21468/SciPostPhysCore.6.4.078}
  {\bibfield  {journal} {\bibinfo  {journal} {SciPost Phys. Core}\ }\textbf
  {\bibinfo {volume} {6}},\ \bibinfo {pages} {078} (\bibinfo {year}
  {2023})}\BibitemShut {NoStop}%
\bibitem [{\citenamefont {Tirrito}\ \emph {et~al.}(2023)\citenamefont
  {Tirrito}, \citenamefont {Santini}, \citenamefont {Fazio},\ and\
  \citenamefont {Collura}}]{tirrito2023fullcountingstatistics}%
  \BibitemOpen
  \bibfield  {author} {\bibinfo {author} {\bibfnamefont {E.}~\bibnamefont
  {Tirrito}}, \bibinfo {author} {\bibfnamefont {A.}~\bibnamefont {Santini}},
  \bibinfo {author} {\bibfnamefont {R.}~\bibnamefont {Fazio}},\ and\ \bibinfo
  {author} {\bibfnamefont {M.}~\bibnamefont {Collura}},\ }\href
  {https://doi.org/10.21468/SciPostPhys.15.3.096} {\bibfield  {journal}
  {\bibinfo  {journal} {SciPost Phys.}\ }\textbf {\bibinfo {volume} {15}},\
  \bibinfo {pages} {096} (\bibinfo {year} {2023})}\BibitemShut {NoStop}%
\bibitem [{\citenamefont {Paviglianiti}\ and\ \citenamefont
  {Silva}(2023)}]{paviglianiti2023multipartiteentanglementin}%
  \BibitemOpen
  \bibfield  {author} {\bibinfo {author} {\bibfnamefont {A.}~\bibnamefont
  {Paviglianiti}}\ and\ \bibinfo {author} {\bibfnamefont {A.}~\bibnamefont
  {Silva}},\ }\href {https://doi.org/10.1103/PhysRevB.108.184302} {\bibfield
  {journal} {\bibinfo  {journal} {Phys. Rev. B}\ }\textbf {\bibinfo {volume}
  {108}},\ \bibinfo {pages} {184302} (\bibinfo {year} {2023})}\BibitemShut
  {NoStop}%
\bibitem [{\citenamefont {Rossini}\ and\ \citenamefont
  {Vicari}(2020)}]{rossini2020measuremendinduceddynamics}%
  \BibitemOpen
  \bibfield  {author} {\bibinfo {author} {\bibfnamefont {D.}~\bibnamefont
  {Rossini}}\ and\ \bibinfo {author} {\bibfnamefont {E.}~\bibnamefont
  {Vicari}},\ }\href {https://doi.org/10.1103/PhysRevB.102.035119} {\bibfield
  {journal} {\bibinfo  {journal} {Phys. Rev. B}\ }\textbf {\bibinfo {volume}
  {102}},\ \bibinfo {pages} {035119} (\bibinfo {year} {2020})}\BibitemShut
  {NoStop}%
\bibitem [{\citenamefont {Tang}\ and\ \citenamefont
  {Zhu}(2020)}]{tang2020measurementinducedphase}%
  \BibitemOpen
  \bibfield  {author} {\bibinfo {author} {\bibfnamefont {Q.}~\bibnamefont
  {Tang}}\ and\ \bibinfo {author} {\bibfnamefont {W.}~\bibnamefont {Zhu}},\
  }\href {https://doi.org/10.1103/PhysRevResearch.2.013022} {\bibfield
  {journal} {\bibinfo  {journal} {Phys. Rev. Res.}\ }\textbf {\bibinfo {volume}
  {2}},\ \bibinfo {pages} {013022} (\bibinfo {year} {2020})}\BibitemShut
  {NoStop}%
\bibitem [{\citenamefont {Fuji}\ and\ \citenamefont
  {Ashida}(2020)}]{fuji2020measurementinducedquantum}%
  \BibitemOpen
  \bibfield  {author} {\bibinfo {author} {\bibfnamefont {Y.}~\bibnamefont
  {Fuji}}\ and\ \bibinfo {author} {\bibfnamefont {Y.}~\bibnamefont {Ashida}},\
  }\href {https://doi.org/10.1103/PhysRevB.102.054302} {\bibfield  {journal}
  {\bibinfo  {journal} {Phys. Rev. B}\ }\textbf {\bibinfo {volume} {102}},\
  \bibinfo {pages} {054302} (\bibinfo {year} {2020})}\BibitemShut {NoStop}%
\bibitem [{\citenamefont {Sierant}\ \emph
  {et~al.}(2022{\natexlab{b}})\citenamefont {Sierant}, \citenamefont
  {Chiriac{\`{o}}}, \citenamefont {Surace}, \citenamefont {Sharma},
  \citenamefont {Turkeshi}, \citenamefont {Dalmonte}, \citenamefont {Fazio},\
  and\ \citenamefont {Pagano}}]{sierant2022dissipativefloquetdynamics}%
  \BibitemOpen
  \bibfield  {author} {\bibinfo {author} {\bibfnamefont {P.}~\bibnamefont
  {Sierant}}, \bibinfo {author} {\bibfnamefont {G.}~\bibnamefont
  {Chiriac{\`{o}}}}, \bibinfo {author} {\bibfnamefont {F.~M.}\ \bibnamefont
  {Surace}}, \bibinfo {author} {\bibfnamefont {S.}~\bibnamefont {Sharma}},
  \bibinfo {author} {\bibfnamefont {X.}~\bibnamefont {Turkeshi}}, \bibinfo
  {author} {\bibfnamefont {M.}~\bibnamefont {Dalmonte}}, \bibinfo {author}
  {\bibfnamefont {R.}~\bibnamefont {Fazio}},\ and\ \bibinfo {author}
  {\bibfnamefont {G.}~\bibnamefont {Pagano}},\ }\href
  {https://doi.org/10.22331/q-2022-02-02-638} {\bibfield  {journal} {\bibinfo
  {journal} {{Quantum}}\ }\textbf {\bibinfo {volume} {6}},\ \bibinfo {pages}
  {638} (\bibinfo {year} {2022}{\natexlab{b}})}\BibitemShut {NoStop}%
\bibitem [{\citenamefont {Doggen}\ \emph {et~al.}(2022)\citenamefont {Doggen},
  \citenamefont {Gefen}, \citenamefont {Gornyi}, \citenamefont {Mirlin},\ and\
  \citenamefont {Polyakov}}]{dogger2022generalizedquantummeasurements}%
  \BibitemOpen
  \bibfield  {author} {\bibinfo {author} {\bibfnamefont {E.~V.~H.}\
  \bibnamefont {Doggen}}, \bibinfo {author} {\bibfnamefont {Y.}~\bibnamefont
  {Gefen}}, \bibinfo {author} {\bibfnamefont {I.~V.}\ \bibnamefont {Gornyi}},
  \bibinfo {author} {\bibfnamefont {A.~D.}\ \bibnamefont {Mirlin}},\ and\
  \bibinfo {author} {\bibfnamefont {D.~G.}\ \bibnamefont {Polyakov}},\ }\href
  {https://doi.org/10.1103/PhysRevResearch.4.023146} {\bibfield  {journal}
  {\bibinfo  {journal} {Phys. Rev. Res.}\ }\textbf {\bibinfo {volume} {4}},\
  \bibinfo {pages} {023146} (\bibinfo {year} {2022})}\BibitemShut {NoStop}%
\bibitem [{\citenamefont {Altland}\ \emph {et~al.}(2022)\citenamefont
  {Altland}, \citenamefont {Buchhold}, \citenamefont {Diehl},\ and\
  \citenamefont {Micklitz}}]{altland2022dynamicsofmeasured}%
  \BibitemOpen
  \bibfield  {author} {\bibinfo {author} {\bibfnamefont {A.}~\bibnamefont
  {Altland}}, \bibinfo {author} {\bibfnamefont {M.}~\bibnamefont {Buchhold}},
  \bibinfo {author} {\bibfnamefont {S.}~\bibnamefont {Diehl}},\ and\ \bibinfo
  {author} {\bibfnamefont {T.}~\bibnamefont {Micklitz}},\ }\href
  {https://doi.org/10.1103/PhysRevResearch.4.L022066} {\bibfield  {journal}
  {\bibinfo  {journal} {Phys. Rev. Res.}\ }\textbf {\bibinfo {volume} {4}},\
  \bibinfo {pages} {L022066} (\bibinfo {year} {2022})}\BibitemShut {NoStop}%
\bibitem [{\citenamefont {Gullans}\ and\ \citenamefont
  {Huse}(2020{\natexlab{a}})}]{gullans2020scalableprobesof}%
  \BibitemOpen
  \bibfield  {author} {\bibinfo {author} {\bibfnamefont {M.~J.}\ \bibnamefont
  {Gullans}}\ and\ \bibinfo {author} {\bibfnamefont {D.~A.}\ \bibnamefont
  {Huse}},\ }\href {https://doi.org/10.1103/PhysRevLett.125.070606} {\bibfield
  {journal} {\bibinfo  {journal} {Phys. Rev. Lett.}\ }\textbf {\bibinfo
  {volume} {125}},\ \bibinfo {pages} {070606} (\bibinfo {year}
  {2020}{\natexlab{a}})}\BibitemShut {NoStop}%
\bibitem [{\citenamefont {Gullans}\ and\ \citenamefont
  {Huse}(2020{\natexlab{b}})}]{gullans2020dynamicalpurificationphase}%
  \BibitemOpen
  \bibfield  {author} {\bibinfo {author} {\bibfnamefont {M.~J.}\ \bibnamefont
  {Gullans}}\ and\ \bibinfo {author} {\bibfnamefont {D.~A.}\ \bibnamefont
  {Huse}},\ }\href {https://doi.org/10.1103/PhysRevX.10.041020} {\bibfield
  {journal} {\bibinfo  {journal} {Phys. Rev. X}\ }\textbf {\bibinfo {volume}
  {10}},\ \bibinfo {pages} {041020} (\bibinfo {year}
  {2020}{\natexlab{b}})}\BibitemShut {NoStop}%
\bibitem [{\citenamefont {L\'oio}\ \emph {et~al.}(2023)\citenamefont {L\'oio},
  \citenamefont {De~Luca}, \citenamefont {De~Nardis},\ and\ \citenamefont
  {Turkeshi}}]{loio2023purificationtimescalesin}%
  \BibitemOpen
  \bibfield  {author} {\bibinfo {author} {\bibfnamefont {H.}~\bibnamefont
  {L\'oio}}, \bibinfo {author} {\bibfnamefont {A.}~\bibnamefont {De~Luca}},
  \bibinfo {author} {\bibfnamefont {J.}~\bibnamefont {De~Nardis}},\ and\
  \bibinfo {author} {\bibfnamefont {X.}~\bibnamefont {Turkeshi}},\ }\href
  {https://doi.org/10.1103/PhysRevB.108.L020306} {\bibfield  {journal}
  {\bibinfo  {journal} {Phys. Rev. B}\ }\textbf {\bibinfo {volume} {108}},\
  \bibinfo {pages} {L020306} (\bibinfo {year} {2023})}\BibitemShut {NoStop}%
\bibitem [{\citenamefont {Choi}\ \emph {et~al.}(2020)\citenamefont {Choi},
  \citenamefont {Bao}, \citenamefont {Qi},\ and\ \citenamefont
  {Altman}}]{choi2020quantumerrorcorrection}%
  \BibitemOpen
  \bibfield  {author} {\bibinfo {author} {\bibfnamefont {S.}~\bibnamefont
  {Choi}}, \bibinfo {author} {\bibfnamefont {Y.}~\bibnamefont {Bao}}, \bibinfo
  {author} {\bibfnamefont {X.-L.}\ \bibnamefont {Qi}},\ and\ \bibinfo {author}
  {\bibfnamefont {E.}~\bibnamefont {Altman}},\ }\href
  {https://doi.org/10.1103/PhysRevLett.125.030505} {\bibfield  {journal}
  {\bibinfo  {journal} {Phys. Rev. Lett.}\ }\textbf {\bibinfo {volume} {125}},\
  \bibinfo {pages} {030505} (\bibinfo {year} {2020})}\BibitemShut {NoStop}%
\bibitem [{\citenamefont {Bao}\ \emph {et~al.}(2020)\citenamefont {Bao},
  \citenamefont {Choi},\ and\ \citenamefont {Altman}}]{bao2020theoryofthe}%
  \BibitemOpen
  \bibfield  {author} {\bibinfo {author} {\bibfnamefont {Y.}~\bibnamefont
  {Bao}}, \bibinfo {author} {\bibfnamefont {S.}~\bibnamefont {Choi}},\ and\
  \bibinfo {author} {\bibfnamefont {E.}~\bibnamefont {Altman}},\ }\href
  {https://doi.org/10.1103/PhysRevB.101.104301} {\bibfield  {journal} {\bibinfo
   {journal} {Phys. Rev. B}\ }\textbf {\bibinfo {volume} {101}},\ \bibinfo
  {pages} {104301} (\bibinfo {year} {2020})}\BibitemShut {NoStop}%
\bibitem [{\citenamefont {Bao}\ \emph {et~al.}(2021)\citenamefont {Bao},
  \citenamefont {Choi},\ and\ \citenamefont
  {Altman}}]{bao2021symmetryenrichedphases}%
  \BibitemOpen
  \bibfield  {author} {\bibinfo {author} {\bibfnamefont {Y.}~\bibnamefont
  {Bao}}, \bibinfo {author} {\bibfnamefont {S.}~\bibnamefont {Choi}},\ and\
  \bibinfo {author} {\bibfnamefont {E.}~\bibnamefont {Altman}},\ }\href
  {https://doi.org/10.1016/j.aop.2021.168618} {\bibfield  {journal} {\bibinfo
  {journal} {Ann. Phys.}\ }\textbf {\bibinfo {volume} {435}},\ \bibinfo {pages}
  {168618} (\bibinfo {year} {2021})}\BibitemShut {NoStop}%
\bibitem [{\citenamefont {Fidkowski}\ \emph {et~al.}(2021)\citenamefont
  {Fidkowski}, \citenamefont {Haah},\ and\ \citenamefont
  {Hastings}}]{fidkowski2021howdynamicalquantum}%
  \BibitemOpen
  \bibfield  {author} {\bibinfo {author} {\bibfnamefont {L.}~\bibnamefont
  {Fidkowski}}, \bibinfo {author} {\bibfnamefont {J.}~\bibnamefont {Haah}},\
  and\ \bibinfo {author} {\bibfnamefont {M.~B.}\ \bibnamefont {Hastings}},\
  }\href {https://doi.org/10.22331/q-2021-01-17-382} {\bibfield  {journal}
  {\bibinfo  {journal} {{Quantum}}\ }\textbf {\bibinfo {volume} {5}},\ \bibinfo
  {pages} {382} (\bibinfo {year} {2021})}\BibitemShut {NoStop}%
\bibitem [{\citenamefont {Bao}\ \emph {et~al.}()\citenamefont {Bao},
  \citenamefont {Block},\ and\ \citenamefont
  {Altman}}]{bao2021finitetimeteleportation}%
  \BibitemOpen
  \bibfield  {author} {\bibinfo {author} {\bibfnamefont {Y.}~\bibnamefont
  {Bao}}, \bibinfo {author} {\bibfnamefont {M.}~\bibnamefont {Block}},\ and\
  \bibinfo {author} {\bibfnamefont {E.}~\bibnamefont {Altman}},\ }\href@noop {}
  {}\Eprint {https://arxiv.org/abs/2110.06963} {arXiv:2110.06963} \BibitemShut
  {NoStop}%
\bibitem [{\citenamefont {Barratt}\ \emph {et~al.}(2022)\citenamefont
  {Barratt}, \citenamefont {Agrawal}, \citenamefont {Potter}, \citenamefont
  {Gopalakrishnan},\ and\ \citenamefont
  {Vasseur}}]{barratt2022transitionsinthe}%
  \BibitemOpen
  \bibfield  {author} {\bibinfo {author} {\bibfnamefont {F.}~\bibnamefont
  {Barratt}}, \bibinfo {author} {\bibfnamefont {U.}~\bibnamefont {Agrawal}},
  \bibinfo {author} {\bibfnamefont {A.~C.}\ \bibnamefont {Potter}}, \bibinfo
  {author} {\bibfnamefont {S.}~\bibnamefont {Gopalakrishnan}},\ and\ \bibinfo
  {author} {\bibfnamefont {R.}~\bibnamefont {Vasseur}},\ }\href
  {https://doi.org/10.1103/PhysRevLett.129.200602} {\bibfield  {journal}
  {\bibinfo  {journal} {Phys. Rev. Lett.}\ }\textbf {\bibinfo {volume} {129}},\
  \bibinfo {pages} {200602} (\bibinfo {year} {2022})}\BibitemShut {NoStop}%
\bibitem [{\citenamefont {Dehghani}\ \emph {et~al.}()\citenamefont {Dehghani},
  \citenamefont {Lavasani}, \citenamefont {Hafezi},\ and\ \citenamefont
  {Gullans}}]{dehgani2022neuralnetworkdecoders}%
  \BibitemOpen
  \bibfield  {author} {\bibinfo {author} {\bibfnamefont {H.}~\bibnamefont
  {Dehghani}}, \bibinfo {author} {\bibfnamefont {A.}~\bibnamefont {Lavasani}},
  \bibinfo {author} {\bibfnamefont {M.}~\bibnamefont {Hafezi}},\ and\ \bibinfo
  {author} {\bibfnamefont {M.~J.}\ \bibnamefont {Gullans}},\ }\href@noop {}
  {}\Eprint {https://arxiv.org/abs/2204.10904} {arXiv:2204.10904} \BibitemShut
  {NoStop}%
\bibitem [{\citenamefont {Kelly}\ \emph {et~al.}(2023)\citenamefont {Kelly},
  \citenamefont {Poschinger}, \citenamefont {Schmidt-Kaler}, \citenamefont
  {Fisher},\ and\ \citenamefont {Marino}}]{kelly2022coherencerequirementsfor}%
  \BibitemOpen
  \bibfield  {author} {\bibinfo {author} {\bibfnamefont {S.~P.}\ \bibnamefont
  {Kelly}}, \bibinfo {author} {\bibfnamefont {U.}~\bibnamefont {Poschinger}},
  \bibinfo {author} {\bibfnamefont {F.}~\bibnamefont {Schmidt-Kaler}}, \bibinfo
  {author} {\bibfnamefont {M.~P.~A.}\ \bibnamefont {Fisher}},\ and\ \bibinfo
  {author} {\bibfnamefont {J.}~\bibnamefont {Marino}},\ }\href
  {https://doi.org/10.21468/SciPostPhys.15.6.250} {\bibfield  {journal}
  {\bibinfo  {journal} {SciPost Phys.}\ }\textbf {\bibinfo {volume} {15}},\
  \bibinfo {pages} {250} (\bibinfo {year} {2023})}\BibitemShut {NoStop}%
\bibitem [{\citenamefont {Noel}\ \emph {et~al.}(2022)\citenamefont {Noel},
  \citenamefont {Niroula}, \citenamefont {Zhu}, \citenamefont {Risinger},
  \citenamefont {Egan}, \citenamefont {Biswas}, \citenamefont {Cetina},
  \citenamefont {Gorshkov}, \citenamefont {Gullans}, \citenamefont {Huse},\
  and\ \citenamefont {Monroe}}]{noel2021measurementinducedquantum}%
  \BibitemOpen
  \bibfield  {author} {\bibinfo {author} {\bibfnamefont {C.}~\bibnamefont
  {Noel}}, \bibinfo {author} {\bibfnamefont {P.}~\bibnamefont {Niroula}},
  \bibinfo {author} {\bibfnamefont {D.}~\bibnamefont {Zhu}}, \bibinfo {author}
  {\bibfnamefont {A.}~\bibnamefont {Risinger}}, \bibinfo {author}
  {\bibfnamefont {L.}~\bibnamefont {Egan}}, \bibinfo {author} {\bibfnamefont
  {D.}~\bibnamefont {Biswas}}, \bibinfo {author} {\bibfnamefont
  {M.}~\bibnamefont {Cetina}}, \bibinfo {author} {\bibfnamefont {A.~V.}\
  \bibnamefont {Gorshkov}}, \bibinfo {author} {\bibfnamefont {M.~J.}\
  \bibnamefont {Gullans}}, \bibinfo {author} {\bibfnamefont {D.~A.}\
  \bibnamefont {Huse}},\ and\ \bibinfo {author} {\bibfnamefont
  {C.}~\bibnamefont {Monroe}},\ }\href
  {https://www.nature.com/articles/s41567-022-01619-7} {\bibfield  {journal}
  {\bibinfo  {journal} {Nature Phys.}\ }\textbf {\bibinfo {volume} {18}},\
  \bibinfo {pages} {760} (\bibinfo {year} {2022})}\BibitemShut {NoStop}%
\bibitem [{\citenamefont {Koh}\ \emph {et~al.}(2023)\citenamefont {Koh},
  \citenamefont {Sun}, \citenamefont {Motta},\ and\ \citenamefont
  {Minnich}}]{koh2022experimentalrealizationof}%
  \BibitemOpen
  \bibfield  {author} {\bibinfo {author} {\bibfnamefont {J.~M.}\ \bibnamefont
  {Koh}}, \bibinfo {author} {\bibfnamefont {S.-N.}\ \bibnamefont {Sun}},
  \bibinfo {author} {\bibfnamefont {M.}~\bibnamefont {Motta}},\ and\ \bibinfo
  {author} {\bibfnamefont {A.~J.}\ \bibnamefont {Minnich}},\ }\href
  {https://doi.org/10.1038/s41567-023-02076-6} {\bibfield  {journal} {\bibinfo
  {journal} {Nature Phys.}\ }\textbf {\bibinfo {volume} {19}},\ \bibinfo
  {pages} {1314} (\bibinfo {year} {2023})}\BibitemShut {NoStop}%
\bibitem [{\citenamefont {Hoke}\ \emph {et~al.}(2023)\citenamefont {Hoke},
  \citenamefont {Ippoliti}, \citenamefont {Rosenberg}, \citenamefont {Abanin},
  \citenamefont {Acharya}, \citenamefont {Andersen}, \citenamefont {Ansmann},
  \citenamefont {Arute}, \citenamefont {Arya}, \citenamefont {Asfaw},
  \citenamefont {Atalaya}, \citenamefont {Bardin}, \citenamefont {Bengtsson},
  \citenamefont {Bortoli}, \citenamefont {Bourassa}, \citenamefont {Bovaird},
  \citenamefont {Brill}, \citenamefont {Broughton}, \citenamefont {Buckley},
  \citenamefont {Buell}, \citenamefont {Burger}, \citenamefont {Burkett},
  \citenamefont {Bushnell}, \citenamefont {Chen}, \citenamefont {Chiaro},
  \citenamefont {Chik}, \citenamefont {Cogan}, \citenamefont {Collins},
  \citenamefont {Conner}, \citenamefont {Courtney}, \citenamefont {Crook},
  \citenamefont {Curtin}, \citenamefont {Dau}, \citenamefont {Debroy},
  \citenamefont {Del Toro~Barba}, \citenamefont {Demura}, \citenamefont
  {Di~Paolo}, \citenamefont {Drozdov}, \citenamefont {Dunsworth}, \citenamefont
  {Eppens}, \citenamefont {Erickson}, \citenamefont {Farhi}, \citenamefont
  {Fatemi}, \citenamefont {Ferreira}, \citenamefont {Burgos}, \citenamefont
  {Forati}, \citenamefont {Fowler}, \citenamefont {Foxen}, \citenamefont
  {Giang}, \citenamefont {Gidney}, \citenamefont {Gilboa}, \citenamefont
  {Giustina}, \citenamefont {Gosula}, \citenamefont {Gross}, \citenamefont
  {Habegger}, \citenamefont {Hamilton}, \citenamefont {Hansen}, \citenamefont
  {Harrigan}, \citenamefont {Harrington}, \citenamefont {Heu}, \citenamefont
  {Hoffmann}, \citenamefont {Hong}, \citenamefont {Huang}, \citenamefont
  {Huff}, \citenamefont {Huggins}, \citenamefont {Isakov}, \citenamefont
  {Iveland}, \citenamefont {Jeffrey}, \citenamefont {Jiang}, \citenamefont
  {Jones}, \citenamefont {Juhas}, \citenamefont {Kafri}, \citenamefont
  {Kechedzhi}, \citenamefont {Khattar}, \citenamefont {Khezri}, \citenamefont
  {Kieferová}, \citenamefont {Kim}, \citenamefont {Kitaev}, \citenamefont
  {Klimov}, \citenamefont {Klots}, \citenamefont {Korotkov}, \citenamefont
  {Kostritsa}, \citenamefont {Kreikebaum}, \citenamefont {Landhuis},
  \citenamefont {Laptev}, \citenamefont {Lau}, \citenamefont {Laws},
  \citenamefont {Lee}, \citenamefont {Lee}, \citenamefont {Lensky},
  \citenamefont {Lester}, \citenamefont {Lill}, \citenamefont {Liu},
  \citenamefont {Locharla}, \citenamefont {Martin}, \citenamefont {McClean},
  \citenamefont {McEwen}, \citenamefont {Miao}, \citenamefont {Mieszala},
  \citenamefont {Montazeri}, \citenamefont {Morvan}, \citenamefont {Movassagh},
  \citenamefont {Mruczkiewicz}, \citenamefont {Neeley}, \citenamefont {Neill},
  \citenamefont {Nersisyan}, \citenamefont {Newman}, \citenamefont {Ng},
  \citenamefont {Nguyen}, \citenamefont {Nguyen}, \citenamefont {Niu},
  \citenamefont {O’Brien}, \citenamefont {Omonije}, \citenamefont {Opremcak},
  \citenamefont {Petukhov}, \citenamefont {Potter}, \citenamefont {Pryadko},
  \citenamefont {Quintana}, \citenamefont {Rocque}, \citenamefont {Rubin},
  \citenamefont {Saei}, \citenamefont {Sank}, \citenamefont {Sankaragomathi},
  \citenamefont {Satzinger}, \citenamefont {Schurkus}, \citenamefont
  {Schuster}, \citenamefont {Shearn}, \citenamefont {Shorter}, \citenamefont
  {Shutty}, \citenamefont {Shvarts}, \citenamefont {Skruzny}, \citenamefont
  {Smith}, \citenamefont {Somma}, \citenamefont {Sterling}, \citenamefont
  {Strain}, \citenamefont {Szalay}, \citenamefont {Torres}, \citenamefont
  {Vidal}, \citenamefont {Villalonga}, \citenamefont {Heidweiller},
  \citenamefont {White}, \citenamefont {Woo}, \citenamefont {Xing},
  \citenamefont {Yao}, \citenamefont {Yeh}, \citenamefont {Yoo}, \citenamefont
  {Young}, \citenamefont {Zalcman}, \citenamefont {Zhang}, \citenamefont {Zhu},
  \citenamefont {Zobrist}, \citenamefont {Neven}, \citenamefont {Babbush},
  \citenamefont {Bacon}, \citenamefont {Boixo}, \citenamefont {Hilton},
  \citenamefont {Lucero}, \citenamefont {Megrant}, \citenamefont {Kelly},
  \citenamefont {Chen}, \citenamefont {Smelyanskiy}, \citenamefont {Mi},
  \citenamefont {Khemani},\ and\ \citenamefont
  {Roushan}}]{hoke2023quantuminformationphases}%
  \BibitemOpen
  \bibfield  {author} {\bibinfo {author} {\bibfnamefont {J.~C.}\ \bibnamefont
  {Hoke}}, \bibinfo {author} {\bibfnamefont {M.}~\bibnamefont {Ippoliti}},
  \bibinfo {author} {\bibfnamefont {E.}~\bibnamefont {Rosenberg}}, \bibinfo
  {author} {\bibfnamefont {D.}~\bibnamefont {Abanin}}, \bibinfo {author}
  {\bibfnamefont {R.}~\bibnamefont {Acharya}}, \bibinfo {author} {\bibfnamefont
  {T.~I.}\ \bibnamefont {Andersen}}, \bibinfo {author} {\bibfnamefont
  {M.}~\bibnamefont {Ansmann}}, \bibinfo {author} {\bibfnamefont
  {F.}~\bibnamefont {Arute}}, \bibinfo {author} {\bibfnamefont
  {K.}~\bibnamefont {Arya}}, \bibinfo {author} {\bibfnamefont {A.}~\bibnamefont
  {Asfaw}}, \bibinfo {author} {\bibfnamefont {J.}~\bibnamefont {Atalaya}},
  \bibinfo {author} {\bibfnamefont {J.~C.}\ \bibnamefont {Bardin}}, \bibinfo
  {author} {\bibfnamefont {A.}~\bibnamefont {Bengtsson}}, \bibinfo {author}
  {\bibfnamefont {G.}~\bibnamefont {Bortoli}}, \bibinfo {author} {\bibfnamefont
  {A.}~\bibnamefont {Bourassa}}, \bibinfo {author} {\bibfnamefont
  {J.}~\bibnamefont {Bovaird}}, \bibinfo {author} {\bibfnamefont
  {L.}~\bibnamefont {Brill}}, \bibinfo {author} {\bibfnamefont
  {M.}~\bibnamefont {Broughton}}, \bibinfo {author} {\bibfnamefont {B.~B.}\
  \bibnamefont {Buckley}}, \bibinfo {author} {\bibfnamefont {D.~A.}\
  \bibnamefont {Buell}}, \bibinfo {author} {\bibfnamefont {T.}~\bibnamefont
  {Burger}}, \bibinfo {author} {\bibfnamefont {B.}~\bibnamefont {Burkett}},
  \bibinfo {author} {\bibfnamefont {N.}~\bibnamefont {Bushnell}}, \bibinfo
  {author} {\bibfnamefont {Z.}~\bibnamefont {Chen}}, \bibinfo {author}
  {\bibfnamefont {B.}~\bibnamefont {Chiaro}}, \bibinfo {author} {\bibfnamefont
  {D.}~\bibnamefont {Chik}}, \bibinfo {author} {\bibfnamefont {J.}~\bibnamefont
  {Cogan}}, \bibinfo {author} {\bibfnamefont {R.}~\bibnamefont {Collins}},
  \bibinfo {author} {\bibfnamefont {P.}~\bibnamefont {Conner}}, \bibinfo
  {author} {\bibfnamefont {W.}~\bibnamefont {Courtney}}, \bibinfo {author}
  {\bibfnamefont {A.~L.}\ \bibnamefont {Crook}}, \bibinfo {author}
  {\bibfnamefont {B.}~\bibnamefont {Curtin}}, \bibinfo {author} {\bibfnamefont
  {A.~G.}\ \bibnamefont {Dau}}, \bibinfo {author} {\bibfnamefont {D.~M.}\
  \bibnamefont {Debroy}}, \bibinfo {author} {\bibfnamefont {A.}~\bibnamefont
  {Del Toro~Barba}}, \bibinfo {author} {\bibfnamefont {S.}~\bibnamefont
  {Demura}}, \bibinfo {author} {\bibfnamefont {A.}~\bibnamefont {Di~Paolo}},
  \bibinfo {author} {\bibfnamefont {I.~K.}\ \bibnamefont {Drozdov}}, \bibinfo
  {author} {\bibfnamefont {A.}~\bibnamefont {Dunsworth}}, \bibinfo {author}
  {\bibfnamefont {D.}~\bibnamefont {Eppens}}, \bibinfo {author} {\bibfnamefont
  {C.}~\bibnamefont {Erickson}}, \bibinfo {author} {\bibfnamefont
  {E.}~\bibnamefont {Farhi}}, \bibinfo {author} {\bibfnamefont
  {R.}~\bibnamefont {Fatemi}}, \bibinfo {author} {\bibfnamefont {V.~S.}\
  \bibnamefont {Ferreira}}, \bibinfo {author} {\bibfnamefont {L.~F.}\
  \bibnamefont {Burgos}}, \bibinfo {author} {\bibfnamefont {E.}~\bibnamefont
  {Forati}}, \bibinfo {author} {\bibfnamefont {A.~G.}\ \bibnamefont {Fowler}},
  \bibinfo {author} {\bibfnamefont {B.}~\bibnamefont {Foxen}}, \bibinfo
  {author} {\bibfnamefont {W.}~\bibnamefont {Giang}}, \bibinfo {author}
  {\bibfnamefont {C.}~\bibnamefont {Gidney}}, \bibinfo {author} {\bibfnamefont
  {D.}~\bibnamefont {Gilboa}}, \bibinfo {author} {\bibfnamefont
  {M.}~\bibnamefont {Giustina}}, \bibinfo {author} {\bibfnamefont
  {R.}~\bibnamefont {Gosula}}, \bibinfo {author} {\bibfnamefont {J.~A.}\
  \bibnamefont {Gross}}, \bibinfo {author} {\bibfnamefont {S.}~\bibnamefont
  {Habegger}}, \bibinfo {author} {\bibfnamefont {M.~C.}\ \bibnamefont
  {Hamilton}}, \bibinfo {author} {\bibfnamefont {M.}~\bibnamefont {Hansen}},
  \bibinfo {author} {\bibfnamefont {M.~P.}\ \bibnamefont {Harrigan}}, \bibinfo
  {author} {\bibfnamefont {S.~D.}\ \bibnamefont {Harrington}}, \bibinfo
  {author} {\bibfnamefont {P.}~\bibnamefont {Heu}}, \bibinfo {author}
  {\bibfnamefont {M.~R.}\ \bibnamefont {Hoffmann}}, \bibinfo {author}
  {\bibfnamefont {S.}~\bibnamefont {Hong}}, \bibinfo {author} {\bibfnamefont
  {T.}~\bibnamefont {Huang}}, \bibinfo {author} {\bibfnamefont
  {A.}~\bibnamefont {Huff}}, \bibinfo {author} {\bibfnamefont {W.~J.}\
  \bibnamefont {Huggins}}, \bibinfo {author} {\bibfnamefont {S.~V.}\
  \bibnamefont {Isakov}}, \bibinfo {author} {\bibfnamefont {J.}~\bibnamefont
  {Iveland}}, \bibinfo {author} {\bibfnamefont {E.}~\bibnamefont {Jeffrey}},
  \bibinfo {author} {\bibfnamefont {Z.}~\bibnamefont {Jiang}}, \bibinfo
  {author} {\bibfnamefont {C.}~\bibnamefont {Jones}}, \bibinfo {author}
  {\bibfnamefont {P.}~\bibnamefont {Juhas}}, \bibinfo {author} {\bibfnamefont
  {D.}~\bibnamefont {Kafri}}, \bibinfo {author} {\bibfnamefont
  {K.}~\bibnamefont {Kechedzhi}}, \bibinfo {author} {\bibfnamefont
  {T.}~\bibnamefont {Khattar}}, \bibinfo {author} {\bibfnamefont
  {M.}~\bibnamefont {Khezri}}, \bibinfo {author} {\bibfnamefont
  {M.}~\bibnamefont {Kieferová}}, \bibinfo {author} {\bibfnamefont
  {S.}~\bibnamefont {Kim}}, \bibinfo {author} {\bibfnamefont {A.}~\bibnamefont
  {Kitaev}}, \bibinfo {author} {\bibfnamefont {P.~V.}\ \bibnamefont {Klimov}},
  \bibinfo {author} {\bibfnamefont {A.~R.}\ \bibnamefont {Klots}}, \bibinfo
  {author} {\bibfnamefont {A.~N.}\ \bibnamefont {Korotkov}}, \bibinfo {author}
  {\bibfnamefont {F.}~\bibnamefont {Kostritsa}}, \bibinfo {author}
  {\bibfnamefont {J.~M.}\ \bibnamefont {Kreikebaum}}, \bibinfo {author}
  {\bibfnamefont {D.}~\bibnamefont {Landhuis}}, \bibinfo {author}
  {\bibfnamefont {P.}~\bibnamefont {Laptev}}, \bibinfo {author} {\bibfnamefont
  {K.-M.}\ \bibnamefont {Lau}}, \bibinfo {author} {\bibfnamefont
  {L.}~\bibnamefont {Laws}}, \bibinfo {author} {\bibfnamefont {J.}~\bibnamefont
  {Lee}}, \bibinfo {author} {\bibfnamefont {K.~W.}\ \bibnamefont {Lee}},
  \bibinfo {author} {\bibfnamefont {Y.~D.}\ \bibnamefont {Lensky}}, \bibinfo
  {author} {\bibfnamefont {B.~J.}\ \bibnamefont {Lester}}, \bibinfo {author}
  {\bibfnamefont {A.~T.}\ \bibnamefont {Lill}}, \bibinfo {author}
  {\bibfnamefont {W.}~\bibnamefont {Liu}}, \bibinfo {author} {\bibfnamefont
  {A.}~\bibnamefont {Locharla}}, \bibinfo {author} {\bibfnamefont
  {O.}~\bibnamefont {Martin}}, \bibinfo {author} {\bibfnamefont {J.~R.}\
  \bibnamefont {McClean}}, \bibinfo {author} {\bibfnamefont {M.}~\bibnamefont
  {McEwen}}, \bibinfo {author} {\bibfnamefont {K.~C.}\ \bibnamefont {Miao}},
  \bibinfo {author} {\bibfnamefont {A.}~\bibnamefont {Mieszala}}, \bibinfo
  {author} {\bibfnamefont {S.}~\bibnamefont {Montazeri}}, \bibinfo {author}
  {\bibfnamefont {A.}~\bibnamefont {Morvan}}, \bibinfo {author} {\bibfnamefont
  {R.}~\bibnamefont {Movassagh}}, \bibinfo {author} {\bibfnamefont
  {W.}~\bibnamefont {Mruczkiewicz}}, \bibinfo {author} {\bibfnamefont
  {M.}~\bibnamefont {Neeley}}, \bibinfo {author} {\bibfnamefont
  {C.}~\bibnamefont {Neill}}, \bibinfo {author} {\bibfnamefont
  {A.}~\bibnamefont {Nersisyan}}, \bibinfo {author} {\bibfnamefont
  {M.}~\bibnamefont {Newman}}, \bibinfo {author} {\bibfnamefont {J.~H.}\
  \bibnamefont {Ng}}, \bibinfo {author} {\bibfnamefont {A.}~\bibnamefont
  {Nguyen}}, \bibinfo {author} {\bibfnamefont {M.}~\bibnamefont {Nguyen}},
  \bibinfo {author} {\bibfnamefont {M.~Y.}\ \bibnamefont {Niu}}, \bibinfo
  {author} {\bibfnamefont {T.~E.}\ \bibnamefont {O’Brien}}, \bibinfo {author}
  {\bibfnamefont {S.}~\bibnamefont {Omonije}}, \bibinfo {author} {\bibfnamefont
  {A.}~\bibnamefont {Opremcak}}, \bibinfo {author} {\bibfnamefont
  {A.}~\bibnamefont {Petukhov}}, \bibinfo {author} {\bibfnamefont
  {R.}~\bibnamefont {Potter}}, \bibinfo {author} {\bibfnamefont {L.~P.}\
  \bibnamefont {Pryadko}}, \bibinfo {author} {\bibfnamefont {C.}~\bibnamefont
  {Quintana}}, \bibinfo {author} {\bibfnamefont {C.}~\bibnamefont {Rocque}},
  \bibinfo {author} {\bibfnamefont {N.~C.}\ \bibnamefont {Rubin}}, \bibinfo
  {author} {\bibfnamefont {N.}~\bibnamefont {Saei}}, \bibinfo {author}
  {\bibfnamefont {D.}~\bibnamefont {Sank}}, \bibinfo {author} {\bibfnamefont
  {K.}~\bibnamefont {Sankaragomathi}}, \bibinfo {author} {\bibfnamefont
  {K.~J.}\ \bibnamefont {Satzinger}}, \bibinfo {author} {\bibfnamefont {H.~F.}\
  \bibnamefont {Schurkus}}, \bibinfo {author} {\bibfnamefont {C.}~\bibnamefont
  {Schuster}}, \bibinfo {author} {\bibfnamefont {M.~J.}\ \bibnamefont
  {Shearn}}, \bibinfo {author} {\bibfnamefont {A.}~\bibnamefont {Shorter}},
  \bibinfo {author} {\bibfnamefont {N.}~\bibnamefont {Shutty}}, \bibinfo
  {author} {\bibfnamefont {V.}~\bibnamefont {Shvarts}}, \bibinfo {author}
  {\bibfnamefont {J.}~\bibnamefont {Skruzny}}, \bibinfo {author} {\bibfnamefont
  {W.~C.}\ \bibnamefont {Smith}}, \bibinfo {author} {\bibfnamefont
  {R.}~\bibnamefont {Somma}}, \bibinfo {author} {\bibfnamefont
  {G.}~\bibnamefont {Sterling}}, \bibinfo {author} {\bibfnamefont
  {D.}~\bibnamefont {Strain}}, \bibinfo {author} {\bibfnamefont
  {M.}~\bibnamefont {Szalay}}, \bibinfo {author} {\bibfnamefont
  {A.}~\bibnamefont {Torres}}, \bibinfo {author} {\bibfnamefont
  {G.}~\bibnamefont {Vidal}}, \bibinfo {author} {\bibfnamefont
  {B.}~\bibnamefont {Villalonga}}, \bibinfo {author} {\bibfnamefont {C.~V.}\
  \bibnamefont {Heidweiller}}, \bibinfo {author} {\bibfnamefont
  {T.}~\bibnamefont {White}}, \bibinfo {author} {\bibfnamefont {B.~W.~K.}\
  \bibnamefont {Woo}}, \bibinfo {author} {\bibfnamefont {C.}~\bibnamefont
  {Xing}}, \bibinfo {author} {\bibfnamefont {Z.~J.}\ \bibnamefont {Yao}},
  \bibinfo {author} {\bibfnamefont {P.}~\bibnamefont {Yeh}}, \bibinfo {author}
  {\bibfnamefont {J.}~\bibnamefont {Yoo}}, \bibinfo {author} {\bibfnamefont
  {G.}~\bibnamefont {Young}}, \bibinfo {author} {\bibfnamefont
  {A.}~\bibnamefont {Zalcman}}, \bibinfo {author} {\bibfnamefont
  {Y.}~\bibnamefont {Zhang}}, \bibinfo {author} {\bibfnamefont
  {N.}~\bibnamefont {Zhu}}, \bibinfo {author} {\bibfnamefont {N.}~\bibnamefont
  {Zobrist}}, \bibinfo {author} {\bibfnamefont {H.}~\bibnamefont {Neven}},
  \bibinfo {author} {\bibfnamefont {R.}~\bibnamefont {Babbush}}, \bibinfo
  {author} {\bibfnamefont {D.}~\bibnamefont {Bacon}}, \bibinfo {author}
  {\bibfnamefont {S.}~\bibnamefont {Boixo}}, \bibinfo {author} {\bibfnamefont
  {J.}~\bibnamefont {Hilton}}, \bibinfo {author} {\bibfnamefont
  {E.}~\bibnamefont {Lucero}}, \bibinfo {author} {\bibfnamefont
  {A.}~\bibnamefont {Megrant}}, \bibinfo {author} {\bibfnamefont
  {J.}~\bibnamefont {Kelly}}, \bibinfo {author} {\bibfnamefont
  {Y.}~\bibnamefont {Chen}}, \bibinfo {author} {\bibfnamefont {V.}~\bibnamefont
  {Smelyanskiy}}, \bibinfo {author} {\bibfnamefont {X.}~\bibnamefont {Mi}},
  \bibinfo {author} {\bibfnamefont {V.}~\bibnamefont {Khemani}},\ and\ \bibinfo
  {author} {\bibfnamefont {P.}~\bibnamefont {Roushan}},\ }\href
  {http://dx.doi.org/10.1038/s41586-023-06505-7} {\bibfield  {journal}
  {\bibinfo  {journal} {Nature}\ }\textbf {\bibinfo {volume} {622}},\ \bibinfo
  {pages} {481} (\bibinfo {year} {2023})}\BibitemShut {NoStop}%
\bibitem [{\citenamefont {Li}\ \emph {et~al.}(2023)\citenamefont {Li},
  \citenamefont {Zou}, \citenamefont {Glorioso}, \citenamefont {Altman},\ and\
  \citenamefont {Fisher}}]{li2022cross}%
  \BibitemOpen
  \bibfield  {author} {\bibinfo {author} {\bibfnamefont {Y.}~\bibnamefont
  {Li}}, \bibinfo {author} {\bibfnamefont {Y.}~\bibnamefont {Zou}}, \bibinfo
  {author} {\bibfnamefont {P.}~\bibnamefont {Glorioso}}, \bibinfo {author}
  {\bibfnamefont {E.}~\bibnamefont {Altman}},\ and\ \bibinfo {author}
  {\bibfnamefont {M.~P.~A.}\ \bibnamefont {Fisher}},\ }\href
  {https://doi.org/10.1103/PhysRevLett.130.220404} {\bibfield  {journal}
  {\bibinfo  {journal} {Phys. Rev. Lett.}\ }\textbf {\bibinfo {volume} {130}},\
  \bibinfo {pages} {220404} (\bibinfo {year} {2023})}\BibitemShut {NoStop}%
\bibitem [{\citenamefont {Li}\ and\ \citenamefont
  {Fisher}(2023)}]{li2021robustdecodingin}%
  \BibitemOpen
  \bibfield  {author} {\bibinfo {author} {\bibfnamefont {Y.}~\bibnamefont
  {Li}}\ and\ \bibinfo {author} {\bibfnamefont {M.~P.~A.}\ \bibnamefont
  {Fisher}},\ }\href {https://doi.org/10.1103/PhysRevB.108.214302} {\bibfield
  {journal} {\bibinfo  {journal} {Phys. Rev. B}\ }\textbf {\bibinfo {volume}
  {108}},\ \bibinfo {pages} {214302} (\bibinfo {year} {2023})}\BibitemShut
  {NoStop}%
\bibitem [{\citenamefont {Garratt}\ and\ \citenamefont
  {Altman}(2023)}]{garratt2023probing}%
  \BibitemOpen
  \bibfield  {author} {\bibinfo {author} {\bibfnamefont {S.~J.}\ \bibnamefont
  {Garratt}}\ and\ \bibinfo {author} {\bibfnamefont {E.}~\bibnamefont
  {Altman}},\ }\href@noop {} {\bibinfo {title} {Probing post-measurement
  entanglement without post-selection}} (\bibinfo {year} {2023}),\ \Eprint
  {https://arxiv.org/abs/2305.20092} {arXiv:2305.20092 [quant-ph]} \BibitemShut
  {NoStop}%
\bibitem [{\citenamefont {Ippoliti}\ and\ \citenamefont
  {Khemani}(2021)}]{ippoliti2021postselectionfreeentanglement}%
  \BibitemOpen
  \bibfield  {author} {\bibinfo {author} {\bibfnamefont {M.}~\bibnamefont
  {Ippoliti}}\ and\ \bibinfo {author} {\bibfnamefont {V.}~\bibnamefont
  {Khemani}},\ }\href {https://doi.org/10.1103/PhysRevLett.126.060501}
  {\bibfield  {journal} {\bibinfo  {journal} {Phys. Rev. Lett.}\ }\textbf
  {\bibinfo {volume} {126}},\ \bibinfo {pages} {060501} (\bibinfo {year}
  {2021})}\BibitemShut {NoStop}%
\bibitem [{\citenamefont {Lu}\ and\ \citenamefont
  {Grover}(2021)}]{lu2021spacetimeduality}%
  \BibitemOpen
  \bibfield  {author} {\bibinfo {author} {\bibfnamefont {T.-C.}\ \bibnamefont
  {Lu}}\ and\ \bibinfo {author} {\bibfnamefont {T.}~\bibnamefont {Grover}},\
  }\href {https://doi.org/10.1103/PRXQuantum.2.040319} {\bibfield  {journal}
  {\bibinfo  {journal} {PRX Quantum}\ }\textbf {\bibinfo {volume} {2}},\
  \bibinfo {pages} {040319} (\bibinfo {year} {2021})}\BibitemShut {NoStop}%
\bibitem [{\citenamefont {Iadecola}\ \emph {et~al.}(2023)\citenamefont
  {Iadecola}, \citenamefont {Ganeshan}, \citenamefont {Pixley},\ and\
  \citenamefont {Wilson}}]{iadecola2022dynamicalentanglementtransition}%
  \BibitemOpen
  \bibfield  {author} {\bibinfo {author} {\bibfnamefont {T.}~\bibnamefont
  {Iadecola}}, \bibinfo {author} {\bibfnamefont {S.}~\bibnamefont {Ganeshan}},
  \bibinfo {author} {\bibfnamefont {J.~H.}\ \bibnamefont {Pixley}},\ and\
  \bibinfo {author} {\bibfnamefont {J.~H.}\ \bibnamefont {Wilson}},\ }\href
  {https://doi.org/10.1103/PhysRevLett.131.060403} {\bibfield  {journal}
  {\bibinfo  {journal} {Phys. Rev. Lett.}\ }\textbf {\bibinfo {volume} {131}},\
  \bibinfo {pages} {060403} (\bibinfo {year} {2023})}\BibitemShut {NoStop}%
\bibitem [{\citenamefont {Buchhold}\ \emph {et~al.}()\citenamefont {Buchhold},
  \citenamefont {M\"uller},\ and\ \citenamefont
  {Diehl}}]{buchhold2022revealingmeasurementinduced}%
  \BibitemOpen
  \bibfield  {author} {\bibinfo {author} {\bibfnamefont {M.}~\bibnamefont
  {Buchhold}}, \bibinfo {author} {\bibfnamefont {T.}~\bibnamefont {M\"uller}},\
  and\ \bibinfo {author} {\bibfnamefont {S.}~\bibnamefont {Diehl}},\
  }\href@noop {} {}\Eprint {https://arxiv.org/abs/2208.10506}
  {arXiv:2208.10506} \BibitemShut {NoStop}%
\bibitem [{\citenamefont {Ravindranath}\ \emph {et~al.}(2023)\citenamefont
  {Ravindranath}, \citenamefont {Han}, \citenamefont {Yang},\ and\
  \citenamefont {Chen}}]{ravindranath2022entanglementsteeringin}%
  \BibitemOpen
  \bibfield  {author} {\bibinfo {author} {\bibfnamefont {V.}~\bibnamefont
  {Ravindranath}}, \bibinfo {author} {\bibfnamefont {Y.}~\bibnamefont {Han}},
  \bibinfo {author} {\bibfnamefont {Z.-C.}\ \bibnamefont {Yang}},\ and\
  \bibinfo {author} {\bibfnamefont {X.}~\bibnamefont {Chen}},\ }\href
  {https://doi.org/10.1103/PhysRevB.108.L041103} {\bibfield  {journal}
  {\bibinfo  {journal} {Phys. Rev. B}\ }\textbf {\bibinfo {volume} {108}},\
  \bibinfo {pages} {L041103} (\bibinfo {year} {2023})}\BibitemShut {NoStop}%
\bibitem [{\citenamefont {O'Dea}\ \emph {et~al.}()\citenamefont {O'Dea},
  \citenamefont {Morningstar}, \citenamefont {Gopalakrishnan},\ and\
  \citenamefont {Khemani}}]{odea2022entanglementandabsorbing}%
  \BibitemOpen
  \bibfield  {author} {\bibinfo {author} {\bibfnamefont {N.}~\bibnamefont
  {O'Dea}}, \bibinfo {author} {\bibfnamefont {A.}~\bibnamefont {Morningstar}},
  \bibinfo {author} {\bibfnamefont {S.}~\bibnamefont {Gopalakrishnan}},\ and\
  \bibinfo {author} {\bibfnamefont {V.}~\bibnamefont {Khemani}},\ }\href@noop
  {} {}\Eprint {https://arxiv.org/abs/2211.12526} {arXiv:2211.12526}
  \BibitemShut {NoStop}%
\bibitem [{\citenamefont {Piroli}\ \emph {et~al.}(2023)\citenamefont {Piroli},
  \citenamefont {Li}, \citenamefont {Vasseur},\ and\ \citenamefont
  {Nahum}}]{piroli2022trivialityofquantum}%
  \BibitemOpen
  \bibfield  {author} {\bibinfo {author} {\bibfnamefont {L.}~\bibnamefont
  {Piroli}}, \bibinfo {author} {\bibfnamefont {Y.}~\bibnamefont {Li}}, \bibinfo
  {author} {\bibfnamefont {R.}~\bibnamefont {Vasseur}},\ and\ \bibinfo {author}
  {\bibfnamefont {A.}~\bibnamefont {Nahum}},\ }\href
  {https://doi.org/10.1103/PhysRevB.107.224303} {\bibfield  {journal} {\bibinfo
   {journal} {Phys. Rev. B}\ }\textbf {\bibinfo {volume} {107}},\ \bibinfo
  {pages} {224303} (\bibinfo {year} {2023})}\BibitemShut {NoStop}%
\bibitem [{\citenamefont {Sierant}\ and\ \citenamefont
  {Turkeshi}(2023)}]{sierant2023controllingentanglementat}%
  \BibitemOpen
  \bibfield  {author} {\bibinfo {author} {\bibfnamefont {P.}~\bibnamefont
  {Sierant}}\ and\ \bibinfo {author} {\bibfnamefont {X.}~\bibnamefont
  {Turkeshi}},\ }\href {https://doi.org/10.1103/PhysRevLett.130.120402}
  {\bibfield  {journal} {\bibinfo  {journal} {Phys. Rev. Lett.}\ }\textbf
  {\bibinfo {volume} {130}},\ \bibinfo {pages} {120402} (\bibinfo {year}
  {2023})}\BibitemShut {NoStop}%
\bibitem [{sup()}]{supplementary}%
  \BibitemOpen
  \href@noop {} {\bibinfo {title} {Supplemental material, where we detail the
  numerical implementations, present additional numerical data, and give a
  qualitative argument for the logarithmic saturation time for the quantum jump
  evolution. it
  includes~\cite{shammah2018openquantumsystems,latorre2005entanglemententropyin,lerose2020originofthe,lerose2020bridgingentanglementdynamics,expokit,gross1982superradiance,daley}}}\BibitemShut
  {NoStop}%
\bibitem [{\citenamefont {Ferioli}\ \emph {et~al.}()\citenamefont {Ferioli},
  \citenamefont {Glicenstein}, \citenamefont {Ferrier-Barbut},\ and\
  \citenamefont {Browaeys}}]{ferioli2023anonequilibrium}%
  \BibitemOpen
  \bibfield  {author} {\bibinfo {author} {\bibfnamefont {G.}~\bibnamefont
  {Ferioli}}, \bibinfo {author} {\bibfnamefont {A.}~\bibnamefont
  {Glicenstein}}, \bibinfo {author} {\bibfnamefont {I.}~\bibnamefont
  {Ferrier-Barbut}},\ and\ \bibinfo {author} {\bibfnamefont {A.}~\bibnamefont
  {Browaeys}},\ }\href@noop {} {}\Eprint {https://arxiv.org/abs/2207.10361}
  {arXiv:2207.10361} \BibitemShut {NoStop}%
\bibitem [{\citenamefont {Iemini}\ \emph {et~al.}(2018)\citenamefont {Iemini},
  \citenamefont {Russomanno}, \citenamefont {Keeling}, \citenamefont
  {Schir\`o}, \citenamefont {Dalmonte},\ and\ \citenamefont
  {Fazio}}]{iemini2018boundarytimecrystals}%
  \BibitemOpen
  \bibfield  {author} {\bibinfo {author} {\bibfnamefont {F.}~\bibnamefont
  {Iemini}}, \bibinfo {author} {\bibfnamefont {A.}~\bibnamefont {Russomanno}},
  \bibinfo {author} {\bibfnamefont {J.}~\bibnamefont {Keeling}}, \bibinfo
  {author} {\bibfnamefont {M.}~\bibnamefont {Schir\`o}}, \bibinfo {author}
  {\bibfnamefont {M.}~\bibnamefont {Dalmonte}},\ and\ \bibinfo {author}
  {\bibfnamefont {R.}~\bibnamefont {Fazio}},\ }\href
  {https://doi.org/10.1103/PhysRevLett.121.035301} {\bibfield  {journal}
  {\bibinfo  {journal} {Phys. Rev. Lett.}\ }\textbf {\bibinfo {volume} {121}},\
  \bibinfo {pages} {035301} (\bibinfo {year} {2018})}\BibitemShut {NoStop}%
\bibitem [{\citenamefont {Hannukainen}\ and\ \citenamefont
  {Larson}(2018)}]{hannukainen2018dissipationdrivenquantum}%
  \BibitemOpen
  \bibfield  {author} {\bibinfo {author} {\bibfnamefont {J.}~\bibnamefont
  {Hannukainen}}\ and\ \bibinfo {author} {\bibfnamefont {J.}~\bibnamefont
  {Larson}},\ }\href {https://doi.org/10.1103/PhysRevA.98.042113} {\bibfield
  {journal} {\bibinfo  {journal} {Phys. Rev. A}\ }\textbf {\bibinfo {volume}
  {98}},\ \bibinfo {pages} {042113} (\bibinfo {year} {2018})}\BibitemShut
  {NoStop}%
\bibitem [{\citenamefont {{Carollo}}\ \emph {et~al.}(2023)\citenamefont
  {{Carollo}}, \citenamefont {{Lesanovsky}}, \citenamefont {{Antezza}},\ and\
  \citenamefont {{De Chiara}}}]{carollo:thermodynamics-btc}%
  \BibitemOpen
  \bibfield  {author} {\bibinfo {author} {\bibfnamefont {F.}~\bibnamefont
  {{Carollo}}}, \bibinfo {author} {\bibfnamefont {I.}~\bibnamefont
  {{Lesanovsky}}}, \bibinfo {author} {\bibfnamefont {M.}~\bibnamefont
  {{Antezza}}},\ and\ \bibinfo {author} {\bibfnamefont {G.}~\bibnamefont {{De
  Chiara}}},\ }\href {https://doi.org/10.48550/arXiv.2306.07330} {\bibfield
  {journal} {\bibinfo  {journal} {arXiv e-prints}\ ,\ \bibinfo {eid}
  {arXiv:2306.07330}} (\bibinfo {year} {2023})},\ \Eprint
  {https://arxiv.org/abs/2306.07330} {arXiv:2306.07330 [quant-ph]} \BibitemShut
  {NoStop}%
\bibitem [{\citenamefont {Cabot}\ \emph {et~al.}()\citenamefont {Cabot},
  \citenamefont {Muhle}, \citenamefont {Carollo},\ and\ \citenamefont
  {Lesanovsky}}]{cabot2022quantum}%
  \BibitemOpen
  \bibfield  {author} {\bibinfo {author} {\bibfnamefont {A.}~\bibnamefont
  {Cabot}}, \bibinfo {author} {\bibfnamefont {L.~S.}\ \bibnamefont {Muhle}},
  \bibinfo {author} {\bibfnamefont {F.}~\bibnamefont {Carollo}},\ and\ \bibinfo
  {author} {\bibfnamefont {I.}~\bibnamefont {Lesanovsky}},\ }\href@noop {}
  {}\Eprint {https://arxiv.org/abs/2212.06460} {arXiv:2212.06460} \BibitemShut
  {NoStop}%
\bibitem [{\citenamefont {Poggi}\ and\ \citenamefont
  {Muñoz-Arias}()}]{poggi2023measurementinducedmmultipartite}%
  \BibitemOpen
  \bibfield  {author} {\bibinfo {author} {\bibfnamefont {P.~M.}\ \bibnamefont
  {Poggi}}\ and\ \bibinfo {author} {\bibfnamefont {M.~H.}\ \bibnamefont
  {Muñoz-Arias}},\ }\href@noop {} {}\Eprint {https://arxiv.org/abs/2305.10209}
  {arXiv:2305.10209} \BibitemShut {NoStop}%
\bibitem [{Note1()}]{Note1}%
  \BibitemOpen
  \bibinfo {note} {This fact hold because the trajectory average does not
  commute with non-linear functions of the quantum trajectory~\cite
  {cao2019entanglementina}.}\BibitemShut {Stop}%
\bibitem [{\citenamefont {{In~\cite{supplementary} we present an example of
  monitored system having a phase transition at the average level, but lacking
  a MIPT.}}()}]{footnote1}%
  \BibitemOpen
  \bibfield  {author} {\bibinfo {author} {\bibnamefont
  {{In~\cite{supplementary} we present an example of monitored system having a
  phase transition at the average level, but lacking a MIPT.}}},\ }\href@noop
  {} {}\BibitemShut {NoStop}%
\bibitem [{\citenamefont {Block}\ \emph {et~al.}(2022)\citenamefont {Block},
  \citenamefont {Bao}, \citenamefont {Choi}, \citenamefont {Altman},\ and\
  \citenamefont {Yao}}]{block2022measurementinducedtransition}%
  \BibitemOpen
  \bibfield  {author} {\bibinfo {author} {\bibfnamefont {M.}~\bibnamefont
  {Block}}, \bibinfo {author} {\bibfnamefont {Y.}~\bibnamefont {Bao}}, \bibinfo
  {author} {\bibfnamefont {S.}~\bibnamefont {Choi}}, \bibinfo {author}
  {\bibfnamefont {E.}~\bibnamefont {Altman}},\ and\ \bibinfo {author}
  {\bibfnamefont {N.~Y.}\ \bibnamefont {Yao}},\ }\href
  {https://doi.org/10.1103/PhysRevLett.128.010604} {\bibfield  {journal}
  {\bibinfo  {journal} {Phys. Rev. Lett.}\ }\textbf {\bibinfo {volume} {128}},\
  \bibinfo {pages} {010604} (\bibinfo {year} {2022})}\BibitemShut {NoStop}%
\bibitem [{\citenamefont {Sharma}\ \emph {et~al.}(2022)\citenamefont {Sharma},
  \citenamefont {Turkeshi}, \citenamefont {Fazio},\ and\ \citenamefont
  {Dalmonte}}]{sharma2022measurementinducedcriticality}%
  \BibitemOpen
  \bibfield  {author} {\bibinfo {author} {\bibfnamefont {S.}~\bibnamefont
  {Sharma}}, \bibinfo {author} {\bibfnamefont {X.}~\bibnamefont {Turkeshi}},
  \bibinfo {author} {\bibfnamefont {R.}~\bibnamefont {Fazio}},\ and\ \bibinfo
  {author} {\bibfnamefont {M.}~\bibnamefont {Dalmonte}},\ }\href
  {https://doi.org/10.21468/SciPostPhysCore.5.2.023} {\bibfield  {journal}
  {\bibinfo  {journal} {SciPost Phys. Core}\ }\textbf {\bibinfo {volume} {5}},\
  \bibinfo {pages} {023} (\bibinfo {year} {2022})}\BibitemShut {NoStop}%
\bibitem [{\citenamefont {Minato}\ \emph {et~al.}(2022)\citenamefont {Minato},
  \citenamefont {Sugimoto}, \citenamefont {Kuwahara},\ and\ \citenamefont
  {Saito}}]{minato2022fateofmeasurementinduced}%
  \BibitemOpen
  \bibfield  {author} {\bibinfo {author} {\bibfnamefont {T.}~\bibnamefont
  {Minato}}, \bibinfo {author} {\bibfnamefont {K.}~\bibnamefont {Sugimoto}},
  \bibinfo {author} {\bibfnamefont {T.}~\bibnamefont {Kuwahara}},\ and\
  \bibinfo {author} {\bibfnamefont {K.}~\bibnamefont {Saito}},\ }\href
  {https://doi.org/10.1103/PhysRevLett.128.010603} {\bibfield  {journal}
  {\bibinfo  {journal} {Phys. Rev. Lett.}\ }\textbf {\bibinfo {volume} {128}},\
  \bibinfo {pages} {010603} (\bibinfo {year} {2022})}\BibitemShut {NoStop}%
\bibitem [{\citenamefont {M\"uller}\ \emph {et~al.}(2022)\citenamefont
  {M\"uller}, \citenamefont {Diehl},\ and\ \citenamefont
  {Buchhold}}]{muller2022measurementinduceddark}%
  \BibitemOpen
  \bibfield  {author} {\bibinfo {author} {\bibfnamefont {T.}~\bibnamefont
  {M\"uller}}, \bibinfo {author} {\bibfnamefont {S.}~\bibnamefont {Diehl}},\
  and\ \bibinfo {author} {\bibfnamefont {M.}~\bibnamefont {Buchhold}},\ }\href
  {https://doi.org/10.1103/PhysRevLett.128.010605} {\bibfield  {journal}
  {\bibinfo  {journal} {Phys. Rev. Lett.}\ }\textbf {\bibinfo {volume} {128}},\
  \bibinfo {pages} {010605} (\bibinfo {year} {2022})}\BibitemShut {NoStop}%
\bibitem [{\citenamefont {Hashizume}\ \emph {et~al.}(2022)\citenamefont
  {Hashizume}, \citenamefont {Bentsen},\ and\ \citenamefont
  {Daley}}]{hashizum2022measurementinducedphase}%
  \BibitemOpen
  \bibfield  {author} {\bibinfo {author} {\bibfnamefont {T.}~\bibnamefont
  {Hashizume}}, \bibinfo {author} {\bibfnamefont {G.}~\bibnamefont {Bentsen}},\
  and\ \bibinfo {author} {\bibfnamefont {A.~J.}\ \bibnamefont {Daley}},\ }\href
  {https://doi.org/10.1103/PhysRevResearch.4.013174} {\bibfield  {journal}
  {\bibinfo  {journal} {Phys. Rev. Res.}\ }\textbf {\bibinfo {volume} {4}},\
  \bibinfo {pages} {013174} (\bibinfo {year} {2022})}\BibitemShut {NoStop}%
\bibitem [{\citenamefont {Zhang}\ \emph {et~al.}(2022)\citenamefont {Zhang},
  \citenamefont {Liu}, \citenamefont {Jian},\ and\ \citenamefont
  {Chen}}]{zhang2022universalentanglementtransitions}%
  \BibitemOpen
  \bibfield  {author} {\bibinfo {author} {\bibfnamefont {P.}~\bibnamefont
  {Zhang}}, \bibinfo {author} {\bibfnamefont {C.}~\bibnamefont {Liu}}, \bibinfo
  {author} {\bibfnamefont {S.-K.}\ \bibnamefont {Jian}},\ and\ \bibinfo
  {author} {\bibfnamefont {X.}~\bibnamefont {Chen}},\ }\href
  {https://doi.org/10.22331/q-2022-05-27-723} {\bibfield  {journal} {\bibinfo
  {journal} {{Quantum}}\ }\textbf {\bibinfo {volume} {6}},\ \bibinfo {pages}
  {723} (\bibinfo {year} {2022})}\BibitemShut {NoStop}%
\bibitem [{\citenamefont {Gross}\ and\ \citenamefont
  {Haroche}(1982)}]{gross1982superradiance}%
  \BibitemOpen
  \bibfield  {author} {\bibinfo {author} {\bibfnamefont {M.}~\bibnamefont
  {Gross}}\ and\ \bibinfo {author} {\bibfnamefont {S.}~\bibnamefont
  {Haroche}},\ }\href
  {https://doi.org/https://doi.org/10.1016/0370-1573(82)90102-8} {\bibfield
  {journal} {\bibinfo  {journal} {Phys. Rep.}\ }\textbf {\bibinfo {volume}
  {93}},\ \bibinfo {pages} {301} (\bibinfo {year} {1982})}\BibitemShut
  {NoStop}%
\bibitem [{\citenamefont {Gisin}\ and\ \citenamefont
  {Percival}(1992)}]{gisin1992thequantumstate}%
  \BibitemOpen
  \bibfield  {author} {\bibinfo {author} {\bibfnamefont {N.}~\bibnamefont
  {Gisin}}\ and\ \bibinfo {author} {\bibfnamefont {I.~C.}\ \bibnamefont
  {Percival}},\ }\href {https://doi.org/10.1088/0305-4470/25/21/023} {\bibfield
   {journal} {\bibinfo  {journal} {J. Phys. A: Math. Theor.}\ }\textbf
  {\bibinfo {volume} {25}},\ \bibinfo {pages} {5677} (\bibinfo {year}
  {1992})}\BibitemShut {NoStop}%
\bibitem [{\citenamefont {Ferioli}()}]{ferioliprivate}%
  \BibitemOpen
  \bibfield  {author} {\bibinfo {author} {\bibfnamefont {G.}~\bibnamefont
  {Ferioli}},\ }\href@noop {} {\bibinfo {title} {Private
  communication.}}\BibitemShut {Stop}%
\bibitem [{\citenamefont {Minoguchi}\ \emph {et~al.}(2022)\citenamefont
  {Minoguchi}, \citenamefont {Rabl},\ and\ \citenamefont
  {Buchhold}}]{minoguchi2022continuousgaussianmeasurements}%
  \BibitemOpen
  \bibfield  {author} {\bibinfo {author} {\bibfnamefont {Y.}~\bibnamefont
  {Minoguchi}}, \bibinfo {author} {\bibfnamefont {P.}~\bibnamefont {Rabl}},\
  and\ \bibinfo {author} {\bibfnamefont {M.}~\bibnamefont {Buchhold}},\ }\href
  {https://doi.org/10.21468/SciPostPhys.12.1.009} {\bibfield  {journal}
  {\bibinfo  {journal} {SciPost Phys.}\ }\textbf {\bibinfo {volume} {12}},\
  \bibinfo {pages} {009} (\bibinfo {year} {2022})}\BibitemShut {NoStop}%
\bibitem [{\citenamefont {Ladewig}\ \emph {et~al.}(2022)\citenamefont
  {Ladewig}, \citenamefont {Diehl},\ and\ \citenamefont
  {Buchhold}}]{ladewig2022monitoredopenfermion}%
  \BibitemOpen
  \bibfield  {author} {\bibinfo {author} {\bibfnamefont {B.}~\bibnamefont
  {Ladewig}}, \bibinfo {author} {\bibfnamefont {S.}~\bibnamefont {Diehl}},\
  and\ \bibinfo {author} {\bibfnamefont {M.}~\bibnamefont {Buchhold}},\ }\href
  {https://doi.org/10.1103/PhysRevResearch.4.033001} {\bibfield  {journal}
  {\bibinfo  {journal} {Phys. Rev. Res.}\ }\textbf {\bibinfo {volume} {4}},\
  \bibinfo {pages} {033001} (\bibinfo {year} {2022})}\BibitemShut {NoStop}%
\bibitem [{\citenamefont {Turkeshi}\ \emph
  {et~al.}(2022{\natexlab{b}})\citenamefont {Turkeshi}, \citenamefont
  {Piroli},\ and\ \citenamefont
  {Schir\'o}}]{turkeshi2022enhancedentanglelmentnegativity}%
  \BibitemOpen
  \bibfield  {author} {\bibinfo {author} {\bibfnamefont {X.}~\bibnamefont
  {Turkeshi}}, \bibinfo {author} {\bibfnamefont {L.}~\bibnamefont {Piroli}},\
  and\ \bibinfo {author} {\bibfnamefont {M.}~\bibnamefont {Schir\'o}},\ }\href
  {https://doi.org/10.1103/PhysRevB.106.024304} {\bibfield  {journal} {\bibinfo
   {journal} {Phys. Rev. B}\ }\textbf {\bibinfo {volume} {106}},\ \bibinfo
  {pages} {024304} (\bibinfo {year} {2022}{\natexlab{b}})}\BibitemShut
  {NoStop}%
\bibitem [{\citenamefont {Weinstein}\ \emph {et~al.}(2022)\citenamefont
  {Weinstein}, \citenamefont {Bao},\ and\ \citenamefont
  {Altman}}]{weinstein2022measurementinducedpower}%
  \BibitemOpen
  \bibfield  {author} {\bibinfo {author} {\bibfnamefont {Z.}~\bibnamefont
  {Weinstein}}, \bibinfo {author} {\bibfnamefont {Y.}~\bibnamefont {Bao}},\
  and\ \bibinfo {author} {\bibfnamefont {E.}~\bibnamefont {Altman}},\ }\href
  {https://doi.org/10.1103/PhysRevLett.129.080501} {\bibfield  {journal}
  {\bibinfo  {journal} {Phys. Rev. Lett.}\ }\textbf {\bibinfo {volume} {129}},\
  \bibinfo {pages} {080501} (\bibinfo {year} {2022})}\BibitemShut {NoStop}%
\bibitem [{\citenamefont {Weinstein}\ \emph {et~al.}(2023)\citenamefont
  {Weinstein}, \citenamefont {Kelly}, \citenamefont {Marino},\ and\
  \citenamefont {Altman}}]{weinstein2022scramblingtransitionin}%
  \BibitemOpen
  \bibfield  {author} {\bibinfo {author} {\bibfnamefont {Z.}~\bibnamefont
  {Weinstein}}, \bibinfo {author} {\bibfnamefont {S.~P.}\ \bibnamefont
  {Kelly}}, \bibinfo {author} {\bibfnamefont {J.}~\bibnamefont {Marino}},\ and\
  \bibinfo {author} {\bibfnamefont {E.}~\bibnamefont {Altman}},\ }\href
  {https://doi.org/10.1103/PhysRevLett.131.220404} {\bibfield  {journal}
  {\bibinfo  {journal} {Phys. Rev. Lett.}\ }\textbf {\bibinfo {volume} {131}},\
  \bibinfo {pages} {220404} (\bibinfo {year} {2023})}\BibitemShut {NoStop}%
\bibitem [{\citenamefont {Hauke}\ \emph {et~al.}(2016)\citenamefont {Hauke},
  \citenamefont {Heyl}, \citenamefont {Tagliacozzo},\ and\ \citenamefont
  {Zoller}}]{Hauke2016}%
  \BibitemOpen
  \bibfield  {author} {\bibinfo {author} {\bibfnamefont {P.}~\bibnamefont
  {Hauke}}, \bibinfo {author} {\bibfnamefont {M.}~\bibnamefont {Heyl}},
  \bibinfo {author} {\bibfnamefont {L.}~\bibnamefont {Tagliacozzo}},\ and\
  \bibinfo {author} {\bibfnamefont {P.}~\bibnamefont {Zoller}},\ }\href
  {https://doi.org/10.1038/nphys3700} {\bibfield  {journal} {\bibinfo
  {journal} {Nature Phys.}\ }\textbf {\bibinfo {volume} {12}},\ \bibinfo
  {pages} {778} (\bibinfo {year} {2016})}\BibitemShut {NoStop}%
\bibitem [{\citenamefont {Pezz\`e}\ \emph {et~al.}(2018)\citenamefont
  {Pezz\`e}, \citenamefont {Smerzi}, \citenamefont {Oberthaler}, \citenamefont
  {Schmied},\ and\ \citenamefont {Treutlein}}]{pezze2018quantummetrologywith}%
  \BibitemOpen
  \bibfield  {author} {\bibinfo {author} {\bibfnamefont {L.}~\bibnamefont
  {Pezz\`e}}, \bibinfo {author} {\bibfnamefont {A.}~\bibnamefont {Smerzi}},
  \bibinfo {author} {\bibfnamefont {M.~K.}\ \bibnamefont {Oberthaler}},
  \bibinfo {author} {\bibfnamefont {R.}~\bibnamefont {Schmied}},\ and\ \bibinfo
  {author} {\bibfnamefont {P.}~\bibnamefont {Treutlein}},\ }\href
  {https://doi.org/10.1103/RevModPhys.90.035005} {\bibfield  {journal}
  {\bibinfo  {journal} {Rev. Mod. Phys.}\ }\textbf {\bibinfo {volume} {90}},\
  \bibinfo {pages} {035005} (\bibinfo {year} {2018})}\BibitemShut {NoStop}%
\bibitem [{\citenamefont {Pappalardi}\ \emph {et~al.}(2017)\citenamefont
  {Pappalardi}, \citenamefont {Russomanno}, \citenamefont {Silva},\ and\
  \citenamefont {Fazio}}]{pappalardi2017multipartiteentanglementafter}%
  \BibitemOpen
  \bibfield  {author} {\bibinfo {author} {\bibfnamefont {S.}~\bibnamefont
  {Pappalardi}}, \bibinfo {author} {\bibfnamefont {A.}~\bibnamefont
  {Russomanno}}, \bibinfo {author} {\bibfnamefont {A.}~\bibnamefont {Silva}},\
  and\ \bibinfo {author} {\bibfnamefont {R.}~\bibnamefont {Fazio}},\ }\href
  {https://doi.org/10.1088/1742-5468/aa6809} {\bibfield  {journal} {\bibinfo
  {journal} {J. Stat. Mech.: Theor. Exp.}\ }\textbf {\bibinfo {volume}
  {2017}},\ \bibinfo {pages} {053104} (\bibinfo {year} {2017})}\BibitemShut
  {NoStop}%
\bibitem [{\citenamefont {Desaules}\ \emph {et~al.}(2022)\citenamefont
  {Desaules}, \citenamefont {Pietracaprina}, \citenamefont {Papi\'{c}},
  \citenamefont {Goold},\ and\ \citenamefont
  {Pappalardi}}]{pappalardi2022extensivemultipartiteentanglement}%
  \BibitemOpen
  \bibfield  {author} {\bibinfo {author} {\bibfnamefont {J.-Y.}\ \bibnamefont
  {Desaules}}, \bibinfo {author} {\bibfnamefont {F.}~\bibnamefont
  {Pietracaprina}}, \bibinfo {author} {\bibfnamefont {Z.}~\bibnamefont
  {Papi\'{c}}}, \bibinfo {author} {\bibfnamefont {J.}~\bibnamefont {Goold}},\
  and\ \bibinfo {author} {\bibfnamefont {S.}~\bibnamefont {Pappalardi}},\
  }\href {https://doi.org/10.1103/PhysRevLett.129.020601} {\bibfield  {journal}
  {\bibinfo  {journal} {Phys. Rev. Lett.}\ }\textbf {\bibinfo {volume} {129}},\
  \bibinfo {pages} {020601} (\bibinfo {year} {2022})}\BibitemShut {NoStop}%
\bibitem [{\citenamefont {Pappalardi}\ \emph {et~al.}(2018)\citenamefont
  {Pappalardi}, \citenamefont {Russomanno}, \citenamefont {\v{Z}unkovi\v{c}},
  \citenamefont {Iemini}, \citenamefont {Silva},\ and\ \citenamefont
  {Fazio}}]{pappalardi2018scramblingandentanglement}%
  \BibitemOpen
  \bibfield  {author} {\bibinfo {author} {\bibfnamefont {S.}~\bibnamefont
  {Pappalardi}}, \bibinfo {author} {\bibfnamefont {A.}~\bibnamefont
  {Russomanno}}, \bibinfo {author} {\bibfnamefont {B.}~\bibnamefont
  {\v{Z}unkovi\v{c}}}, \bibinfo {author} {\bibfnamefont {F.}~\bibnamefont
  {Iemini}}, \bibinfo {author} {\bibfnamefont {A.}~\bibnamefont {Silva}},\ and\
  \bibinfo {author} {\bibfnamefont {R.}~\bibnamefont {Fazio}},\ }\href
  {https://doi.org/10.1103/PhysRevB.98.134303} {\bibfield  {journal} {\bibinfo
  {journal} {Phys. Rev. B}\ }\textbf {\bibinfo {volume} {98}},\ \bibinfo
  {pages} {134303} (\bibinfo {year} {2018})}\BibitemShut {NoStop}%
\bibitem [{\citenamefont {Brenes}\ \emph {et~al.}(2020)\citenamefont {Brenes},
  \citenamefont {Pappalardi}, \citenamefont {Goold},\ and\ \citenamefont
  {Silva}}]{brenes2020multipartitenetanglementstructure}%
  \BibitemOpen
  \bibfield  {author} {\bibinfo {author} {\bibfnamefont {M.}~\bibnamefont
  {Brenes}}, \bibinfo {author} {\bibfnamefont {S.}~\bibnamefont {Pappalardi}},
  \bibinfo {author} {\bibfnamefont {J.}~\bibnamefont {Goold}},\ and\ \bibinfo
  {author} {\bibfnamefont {A.}~\bibnamefont {Silva}},\ }\href
  {https://doi.org/10.1103/PhysRevLett.124.040605} {\bibfield  {journal}
  {\bibinfo  {journal} {Phys. Rev. Lett.}\ }\textbf {\bibinfo {volume} {124}},\
  \bibinfo {pages} {040605} (\bibinfo {year} {2020})}\BibitemShut {NoStop}%
\bibitem [{\citenamefont {Dooley}\ \emph {et~al.}(2023)\citenamefont {Dooley},
  \citenamefont {Pappalardi},\ and\ \citenamefont
  {Goold}}]{dooley2023entanglementenhanced}%
  \BibitemOpen
  \bibfield  {author} {\bibinfo {author} {\bibfnamefont {S.}~\bibnamefont
  {Dooley}}, \bibinfo {author} {\bibfnamefont {S.}~\bibnamefont {Pappalardi}},\
  and\ \bibinfo {author} {\bibfnamefont {J.}~\bibnamefont {Goold}},\ }\href
  {https://doi.org/10.1103/PhysRevB.107.035123} {\bibfield  {journal} {\bibinfo
   {journal} {Phys. Rev. B}\ }\textbf {\bibinfo {volume} {107}},\ \bibinfo
  {pages} {035123} (\bibinfo {year} {2023})}\BibitemShut {NoStop}%
\bibitem [{Note2()}]{Note2}%
  \BibitemOpen
  \bibinfo {note} {\protect \textit {En passant}, we note that for a pure state
  $\rho _\protect \mathrm {w}=|\Psi \rangle _\protect \mathrm {w}\langle \Psi
  |_\protect \mathrm {w}$, $F_Q(\rho _\protect \mathrm {w})$ is the maximal
  eigenvalue of the covariance matrix $M^\protect \mathrm {cov}_{\alpha ,\beta
  }=2 \langle \Psi _\protect \mathrm {w} |\{ \protect \hat {J}_\alpha ,\protect
  \hat {J}_\beta \} |\Psi \rangle - 4\langle \Psi _\protect \mathrm {w}
  |\protect \hat {J}_\alpha |\Psi _\protect \mathrm {w}\rangle \langle \Psi
  _\protect \mathrm {w} |\protect \hat {J}_\beta |\Psi _\protect \mathrm
  {w}\rangle $.}\BibitemShut {Stop}%
\bibitem [{Note3()}]{Note3}%
  \BibitemOpen
  \bibinfo {note} {We note that in the quantum jump case, the Fisher density is
  extensive in the system size $\protect \overline {f_Q}\sim N$, cf.~Ref.~\cite
  {supplementary}}\BibitemShut {NoStop}%
\bibitem [{Note4()}]{Note4}%
  \BibitemOpen
  \bibinfo {note} {This fact reveals enhanced multipartite entanglement at the
  critical point}\BibitemShut {NoStop}%
\bibitem [{\citenamefont {Louren{\c{c}}o}\ \emph {et~al.}(2022)\citenamefont
  {Louren{\c{c}}o}, \citenamefont {Prazeres}, \citenamefont {Maciel},
  \citenamefont {Iemini},\ and\ \citenamefont {Duzzioni}}]{fernando}%
  \BibitemOpen
  \bibfield  {author} {\bibinfo {author} {\bibfnamefont {A.~C.}\ \bibnamefont
  {Louren{\c{c}}o}}, \bibinfo {author} {\bibfnamefont {L.~F.~d.}\ \bibnamefont
  {Prazeres}}, \bibinfo {author} {\bibfnamefont {T.~O.}\ \bibnamefont
  {Maciel}}, \bibinfo {author} {\bibfnamefont {F.}~\bibnamefont {Iemini}},\
  and\ \bibinfo {author} {\bibfnamefont {E.~I.}\ \bibnamefont {Duzzioni}},\
  }\href {https://doi.org/10.1103/PhysRevB.105.134422} {\bibfield  {journal}
  {\bibinfo  {journal} {Phys. Rev. B}\ }\textbf {\bibinfo {volume} {105}},\
  \bibinfo {pages} {134422} (\bibinfo {year} {2022})}\BibitemShut {NoStop}%
\bibitem [{\citenamefont {{F. Iemini and S. Pal}}()}]{fernando2}%
  \BibitemOpen
  \bibfield  {author} {\bibinfo {author} {\bibnamefont {{F. Iemini and S.
  Pal}}},\ }\href@noop {} {\bibinfo {title} {Unpublished}}\BibitemShut
  {NoStop}%
\bibitem [{\citenamefont {Ferioli}\ \emph {et~al.}(2021)\citenamefont
  {Ferioli}, \citenamefont {Glicenstein}, \citenamefont {Robicheaux},
  \citenamefont {Sutherland}, \citenamefont {Browaeys},\ and\ \citenamefont
  {Ferrier-Barbut}}]{ferioli2021laser}%
  \BibitemOpen
  \bibfield  {author} {\bibinfo {author} {\bibfnamefont {G.}~\bibnamefont
  {Ferioli}}, \bibinfo {author} {\bibfnamefont {A.}~\bibnamefont
  {Glicenstein}}, \bibinfo {author} {\bibfnamefont {F.}~\bibnamefont
  {Robicheaux}}, \bibinfo {author} {\bibfnamefont {R.~T.}\ \bibnamefont
  {Sutherland}}, \bibinfo {author} {\bibfnamefont {A.}~\bibnamefont
  {Browaeys}},\ and\ \bibinfo {author} {\bibfnamefont {I.}~\bibnamefont
  {Ferrier-Barbut}},\ }\href {https://doi.org/10.1103/PhysRevLett.127.243602}
  {\bibfield  {journal} {\bibinfo  {journal} {Phys. Rev. Lett.}\ }\textbf
  {\bibinfo {volume} {127}},\ \bibinfo {pages} {243602} (\bibinfo {year}
  {2021})}\BibitemShut {NoStop}%
\bibitem [{\citenamefont {Roch}\ \emph {et~al.}(2014)\citenamefont {Roch},
  \citenamefont {Schwartz}, \citenamefont {Motzoi}, \citenamefont {Macklin},
  \citenamefont {Vijay}, \citenamefont {Eddins}, \citenamefont {Korotkov},
  \citenamefont {Whaley}, \citenamefont {Sarovar},\ and\ \citenamefont
  {Siddiqi}}]{roch2014observation}%
  \BibitemOpen
  \bibfield  {author} {\bibinfo {author} {\bibfnamefont {N.}~\bibnamefont
  {Roch}}, \bibinfo {author} {\bibfnamefont {M.~E.}\ \bibnamefont {Schwartz}},
  \bibinfo {author} {\bibfnamefont {F.}~\bibnamefont {Motzoi}}, \bibinfo
  {author} {\bibfnamefont {C.}~\bibnamefont {Macklin}}, \bibinfo {author}
  {\bibfnamefont {R.}~\bibnamefont {Vijay}}, \bibinfo {author} {\bibfnamefont
  {A.~W.}\ \bibnamefont {Eddins}}, \bibinfo {author} {\bibfnamefont {A.~N.}\
  \bibnamefont {Korotkov}}, \bibinfo {author} {\bibfnamefont {K.~B.}\
  \bibnamefont {Whaley}}, \bibinfo {author} {\bibfnamefont {M.}~\bibnamefont
  {Sarovar}},\ and\ \bibinfo {author} {\bibfnamefont {I.}~\bibnamefont
  {Siddiqi}},\ }\href {https://doi.org/10.1103/PhysRevLett.112.170501}
  {\bibfield  {journal} {\bibinfo  {journal} {Phys. Rev. Lett.}\ }\textbf
  {\bibinfo {volume} {112}},\ \bibinfo {pages} {170501} (\bibinfo {year}
  {2014})}\BibitemShut {NoStop}%
\bibitem [{\citenamefont {Campagne-Ibarcq}\ \emph {et~al.}(2016)\citenamefont
  {Campagne-Ibarcq}, \citenamefont {Six}, \citenamefont {Bretheau},
  \citenamefont {Sarlette}, \citenamefont {Mirrahimi}, \citenamefont
  {Rouchon},\ and\ \citenamefont {Huard}}]{campagne2016observing}%
  \BibitemOpen
  \bibfield  {author} {\bibinfo {author} {\bibfnamefont {P.}~\bibnamefont
  {Campagne-Ibarcq}}, \bibinfo {author} {\bibfnamefont {P.}~\bibnamefont
  {Six}}, \bibinfo {author} {\bibfnamefont {L.}~\bibnamefont {Bretheau}},
  \bibinfo {author} {\bibfnamefont {A.}~\bibnamefont {Sarlette}}, \bibinfo
  {author} {\bibfnamefont {M.}~\bibnamefont {Mirrahimi}}, \bibinfo {author}
  {\bibfnamefont {P.}~\bibnamefont {Rouchon}},\ and\ \bibinfo {author}
  {\bibfnamefont {B.}~\bibnamefont {Huard}},\ }\href
  {https://doi.org/10.1103/PhysRevX.6.011002} {\bibfield  {journal} {\bibinfo
  {journal} {Phys. Rev. X}\ }\textbf {\bibinfo {volume} {6}},\ \bibinfo {pages}
  {011002} (\bibinfo {year} {2016})}\BibitemShut {NoStop}%
\bibitem [{Note5()}]{Note5}%
  \BibitemOpen
  \bibinfo {note} {Nevertheless, no direct relationship is in general present
  between these variances and the entanglement entropy.}\BibitemShut {Stop}%
\bibitem [{\citenamefont {Ke\ss{}ler}\ \emph {et~al.}(2021)\citenamefont
  {Ke\ss{}ler}, \citenamefont {Kongkhambut}, \citenamefont {Georges},
  \citenamefont {Mathey}, \citenamefont {Cosme},\ and\ \citenamefont
  {Hemmerich}}]{kessler2021observation}%
  \BibitemOpen
  \bibfield  {author} {\bibinfo {author} {\bibfnamefont {H.}~\bibnamefont
  {Ke\ss{}ler}}, \bibinfo {author} {\bibfnamefont {P.}~\bibnamefont
  {Kongkhambut}}, \bibinfo {author} {\bibfnamefont {C.}~\bibnamefont
  {Georges}}, \bibinfo {author} {\bibfnamefont {L.}~\bibnamefont {Mathey}},
  \bibinfo {author} {\bibfnamefont {J.~G.}\ \bibnamefont {Cosme}},\ and\
  \bibinfo {author} {\bibfnamefont {A.}~\bibnamefont {Hemmerich}},\ }\href
  {https://doi.org/10.1103/PhysRevLett.127.043602} {\bibfield  {journal}
  {\bibinfo  {journal} {Phys. Rev. Lett.}\ }\textbf {\bibinfo {volume} {127}},\
  \bibinfo {pages} {043602} (\bibinfo {year} {2021})}\BibitemShut {NoStop}%
\bibitem [{\citenamefont {Wang}\ \emph {et~al.}(2020)\citenamefont {Wang},
  \citenamefont {Li}, \citenamefont {Feng}, \citenamefont {Song}, \citenamefont
  {Song}, \citenamefont {Liu}, \citenamefont {Guo}, \citenamefont {Zhang},
  \citenamefont {Dong}, \citenamefont {Zheng}, \citenamefont {Wang},\ and\
  \citenamefont {Wang}}]{wang2020controllable}%
  \BibitemOpen
  \bibfield  {author} {\bibinfo {author} {\bibfnamefont {Z.}~\bibnamefont
  {Wang}}, \bibinfo {author} {\bibfnamefont {H.}~\bibnamefont {Li}}, \bibinfo
  {author} {\bibfnamefont {W.}~\bibnamefont {Feng}}, \bibinfo {author}
  {\bibfnamefont {X.}~\bibnamefont {Song}}, \bibinfo {author} {\bibfnamefont
  {C.}~\bibnamefont {Song}}, \bibinfo {author} {\bibfnamefont {W.}~\bibnamefont
  {Liu}}, \bibinfo {author} {\bibfnamefont {Q.}~\bibnamefont {Guo}}, \bibinfo
  {author} {\bibfnamefont {X.}~\bibnamefont {Zhang}}, \bibinfo {author}
  {\bibfnamefont {H.}~\bibnamefont {Dong}}, \bibinfo {author} {\bibfnamefont
  {D.}~\bibnamefont {Zheng}}, \bibinfo {author} {\bibfnamefont
  {H.}~\bibnamefont {Wang}},\ and\ \bibinfo {author} {\bibfnamefont {D.-W.}\
  \bibnamefont {Wang}},\ }\href
  {https://doi.org/10.1103/PhysRevLett.124.013601} {\bibfield  {journal}
  {\bibinfo  {journal} {Phys. Rev. Lett.}\ }\textbf {\bibinfo {volume} {124}},\
  \bibinfo {pages} {013601} (\bibinfo {year} {2020})}\BibitemShut {NoStop}%
\bibitem [{\citenamefont {Liedl}\ \emph {et~al.}(2022)\citenamefont {Liedl},
  \citenamefont {Tebbenjohanns}, \citenamefont {Bach}, \citenamefont {Pucher},
  \citenamefont {Rauschenbeutel},\ and\ \citenamefont
  {Schneeweiss}}]{liedl2022observation}%
  \BibitemOpen
  \bibfield  {author} {\bibinfo {author} {\bibfnamefont {C.}~\bibnamefont
  {Liedl}}, \bibinfo {author} {\bibfnamefont {F.}~\bibnamefont
  {Tebbenjohanns}}, \bibinfo {author} {\bibfnamefont {C.}~\bibnamefont {Bach}},
  \bibinfo {author} {\bibfnamefont {S.}~\bibnamefont {Pucher}}, \bibinfo
  {author} {\bibfnamefont {A.}~\bibnamefont {Rauschenbeutel}},\ and\ \bibinfo
  {author} {\bibfnamefont {P.}~\bibnamefont {Schneeweiss}},\ }\href@noop {}
  {\bibinfo {title} {Observation of superradiant bursts in waveguide qed}}
  (\bibinfo {year} {2022}),\ \Eprint {https://arxiv.org/abs/2211.08940}
  {arXiv:2211.08940 [quant-ph]} \BibitemShut {NoStop}%
\bibitem [{\citenamefont {Zhu}\ \emph {et~al.}(2019)\citenamefont {Zhu},
  \citenamefont {Marino}, \citenamefont {Yao}, \citenamefont {Lukin},\ and\
  \citenamefont {Demler}}]{Zhu_2019}%
  \BibitemOpen
  \bibfield  {author} {\bibinfo {author} {\bibfnamefont {B.}~\bibnamefont
  {Zhu}}, \bibinfo {author} {\bibfnamefont {J.}~\bibnamefont {Marino}},
  \bibinfo {author} {\bibfnamefont {N.~Y.}\ \bibnamefont {Yao}}, \bibinfo
  {author} {\bibfnamefont {M.~D.}\ \bibnamefont {Lukin}},\ and\ \bibinfo
  {author} {\bibfnamefont {E.~A.}\ \bibnamefont {Demler}},\ }\href
  {https://doi.org/10.1088/1367-2630/ab2afe} {\bibfield  {journal} {\bibinfo
  {journal} {New Journal of Physics}\ }\textbf {\bibinfo {volume} {21}},\
  \bibinfo {pages} {073028} (\bibinfo {year} {2019})}\BibitemShut {NoStop}%
\bibitem [{\citenamefont {Shammah}\ \emph {et~al.}(2018)\citenamefont
  {Shammah}, \citenamefont {Ahmed}, \citenamefont {Lambert}, \citenamefont
  {De~Liberato},\ and\ \citenamefont {Nori}}]{shammah2018openquantumsystems}%
  \BibitemOpen
  \bibfield  {author} {\bibinfo {author} {\bibfnamefont {N.}~\bibnamefont
  {Shammah}}, \bibinfo {author} {\bibfnamefont {S.}~\bibnamefont {Ahmed}},
  \bibinfo {author} {\bibfnamefont {N.}~\bibnamefont {Lambert}}, \bibinfo
  {author} {\bibfnamefont {S.}~\bibnamefont {De~Liberato}},\ and\ \bibinfo
  {author} {\bibfnamefont {F.}~\bibnamefont {Nori}},\ }\href
  {https://doi.org/10.1103/PhysRevA.98.063815} {\bibfield  {journal} {\bibinfo
  {journal} {Phys. Rev. A}\ }\textbf {\bibinfo {volume} {98}},\ \bibinfo
  {pages} {063815} (\bibinfo {year} {2018})}\BibitemShut {NoStop}%
\bibitem [{\citenamefont {Latorre}\ \emph {et~al.}(2005)\citenamefont
  {Latorre}, \citenamefont {Or\'us}, \citenamefont {Rico},\ and\ \citenamefont
  {Vidal}}]{latorre2005entanglemententropyin}%
  \BibitemOpen
  \bibfield  {author} {\bibinfo {author} {\bibfnamefont {J.~I.}\ \bibnamefont
  {Latorre}}, \bibinfo {author} {\bibfnamefont {R.}~\bibnamefont {Or\'us}},
  \bibinfo {author} {\bibfnamefont {E.}~\bibnamefont {Rico}},\ and\ \bibinfo
  {author} {\bibfnamefont {J.}~\bibnamefont {Vidal}},\ }\href
  {https://doi.org/10.1103/PhysRevA.71.064101} {\bibfield  {journal} {\bibinfo
  {journal} {Phys. Rev. A}\ }\textbf {\bibinfo {volume} {71}},\ \bibinfo
  {pages} {064101} (\bibinfo {year} {2005})}\BibitemShut {NoStop}%
\bibitem [{\citenamefont {Lerose}\ and\ \citenamefont
  {Pappalardi}(2020{\natexlab{a}})}]{lerose2020originofthe}%
  \BibitemOpen
  \bibfield  {author} {\bibinfo {author} {\bibfnamefont {A.}~\bibnamefont
  {Lerose}}\ and\ \bibinfo {author} {\bibfnamefont {S.}~\bibnamefont
  {Pappalardi}},\ }\href {https://doi.org/10.1103/PhysRevResearch.2.012041}
  {\bibfield  {journal} {\bibinfo  {journal} {Phys. Rev. Res.}\ }\textbf
  {\bibinfo {volume} {2}},\ \bibinfo {pages} {012041} (\bibinfo {year}
  {2020}{\natexlab{a}})}\BibitemShut {NoStop}%
\bibitem [{\citenamefont {Lerose}\ and\ \citenamefont
  {Pappalardi}(2020{\natexlab{b}})}]{lerose2020bridgingentanglementdynamics}%
  \BibitemOpen
  \bibfield  {author} {\bibinfo {author} {\bibfnamefont {A.}~\bibnamefont
  {Lerose}}\ and\ \bibinfo {author} {\bibfnamefont {S.}~\bibnamefont
  {Pappalardi}},\ }\href {https://doi.org/10.1103/PhysRevA.102.032404}
  {\bibfield  {journal} {\bibinfo  {journal} {Phys. Rev. A}\ }\textbf {\bibinfo
  {volume} {102}},\ \bibinfo {pages} {032404} (\bibinfo {year}
  {2020}{\natexlab{b}})}\BibitemShut {NoStop}%
\bibitem [{\citenamefont {Sidje}(1998)}]{expokit}%
  \BibitemOpen
  \bibfield  {author} {\bibinfo {author} {\bibfnamefont {R.~B.}\ \bibnamefont
  {Sidje}},\ }\href@noop {} {\bibfield  {journal} {\bibinfo  {journal} {ACM
  Trans. Math. Softw.}\ }\textbf {\bibinfo {volume} {24}},\ \bibinfo {pages}
  {130} (\bibinfo {year} {1998})}\BibitemShut {NoStop}%
\bibitem [{\citenamefont {Daley}(2014)}]{daley}%
  \BibitemOpen
  \bibfield  {author} {\bibinfo {author} {\bibfnamefont {A.~J.}\ \bibnamefont
  {Daley}},\ }\href {https://doi.org/10.1080%2F00018732.2014.933502} {\bibfield
   {journal} {\bibinfo  {journal} {Adv. Phys.}\ }\textbf {\bibinfo {volume}
  {63}},\ \bibinfo {pages} {77} (\bibinfo {year} {2014})}\BibitemShut {NoStop}%
\end{thebibliography}


\end{document}